%% file: main.tex
\documentclass[a4paper,11pt]{article}
\usepackage{jheppub}

\usepackage{amsmath}
\usepackage{mathtools}
\usepackage{multirow}
\usepackage{array}

\usepackage{graphicx}
\usepackage{lscape}
\usepackage{subfig}

\usepackage[usenames,dvipsnames,svgnames,table]{xcolor}
\usepackage[normalem]{ulem}

\DeclareMathOperator{\shuffle}{\sqcup\mathchoice{\mkern-2.9mu}{\mkern-2.9mu}{\mkern-3mu}{\mkern-3.5mu}\sqcup}

\newcommand{\bea}{\begin{eqnarray}}
\newcommand{\eea}{\end{eqnarray}}
\newcommand{\bei}{\begin{itemize}}
\newcommand{\eei}{\end{itemize}}
\newcommand{\bean}{\begin{eqnarray*}}
\newcommand{\eean}{\end{eqnarray*}}
\newcommand{\nn}{\nonumber \\}

\newcommand{\der}[1]{\operatorname{\partial}_{#1}}

\def\eps{\epsilon}
\def\e{\epsilon}
\def\magnus #1{\Omega [ #1]}

\def\top #1{\mathcal{T}_{#1}}

\newcommand\scalemath[2]{\scalebox{#1}{\mbox{\ensuremath{\displaystyle #2}}}}

\title{Three-loop master integrals for ladder-box diagrams with one massive leg}

\author[a]{Stefano Di Vita,}
\author[a,b]{Pierpaolo Mastrolia,}
\author[a]{Ulrich Schubert,}
\author[a]{and Valery Yundin}

\emailAdd{divita@mpp.mpg.de}
\emailAdd{pierpaolo.mastrolia@cern.ch}
\emailAdd{schubert@mpp.mpg.de}
\emailAdd{yundin@mpp.mpg.de}

\affiliation[a]{Max-Planck-Institut f\"ur Physik, \\ F\"ohringer Ring 6, D-80805 M\"unchen, Germany}
\affiliation[b]{Dipartimento di Fisica e Astronomia, Universit\`a di Padova, and INFN Sezione di Padova, \\ Via Marzolo 8, I-35131 Padova, Italy}

\arxivnumber{1408.xxxx}

\preprint{MPP-2014-324}


\abstract{
The three-loop master integrals for ladder-box diagrams with one massive leg
are computed from an eighty-five by eighty-five system of differential equations, 
solved by means of Magnus exponential.
The results of the considered box-type integrals, as well as of the
tower of vertex- and bubble-type master integrals associated to
subtopologies, are given as
a Taylor series expansion in the dimensional regulator parameter
$\epsilon = (4-d)/2$.
The coefficients of the series are expressed in terms of uniform weight
combinations of multiple polylogarithms and transcendental constants 
up to weight six. 
The considered integrals enter the next-to-next-to-next-to-leading order virtual corrections 
to scattering processes like 
the three-jet production mediated by vector boson decay, $V^{*} \to jjj$, 
as well as the Higgs plus one-jet production in gluon fusion, $pp \to Hj$.
}

\begin{document}

\maketitle
\flushbottom

\input{intro}

\input{diffeq}

\input{2loop}

\input{3loop}
\input{boundaries}

\input{conclusions}

\section*{Acknowledgments}
We whish to thank 
Thomas Gehrmann and Lorenzo Tancredi for discussions on multiple polylogarithms and comparisons.
We also acknowledge Thomas Gehrmann and Ettore Remiddi for comments on
the manuscript. 
P.M. acknowledges the kind hospitality of the CERN Theory Department
during the completion of this work.
The work of  P.M., U.S. and V.Y. is supported by the Alexander von Humboldt
Foundation, in the framework of the Sofja Kovalevskaja Award 2010, endowed by the German Federal Ministry of Education and Research.
This research used resources of the National Energy Research Scientific Computing Center,
 a DOE Office of Science User Facility supported by the Office of Science
 of the U.S. Department of Energy under Contract No.~DE-AC02-05CH11231.

\appendix
\input{2loopMatrices}
\input{3loopMatrices}

\input{3loopg83}

\bibliographystyle{JHEP}
\bibliography{references}

\end{document}

%% file: intro.tex
\section{Introduction}
\label{sec:intro}

Feynman integrals are the building blocks for describing particle
interactions beyond the tree approximation in perturbation theory.
When their direct integration becomes prohibitive, because of either
the number of loops or of external legs, the evaluation of
Feynman integrals can be addressed in two phases, namely the
{\it decomposition} in terms of basic integrals, followed by the 
{\it evaluation} of the latter, to be considered as an easier problem.

Within the continuous dimensional
regularization scheme, Feynman integrals fulfill identities
that fall in the category of the general class of
integration-by-parts relations
\cite{Chetyrkin:1980pr,Tkachov:1981wb,Lee:2008tj}. Such relations can be exploited
in order to {\it identify} a set of independent integrals, 
dubbed {\it master integrals} (MI's),  that can be used as a basis of functions for the 
virtual contributions to scattering amplitudes. 
Therefore, for any given process,
it is possible to identify a minimal set of MI's,
such that {\it all}
Feynman integrals contributing to it admit a representation in terms
of the chosen basis of MI's.

Scattering amplitudes are combinations of Feynman diagrams. 
Therefore, it is not unnatural that global properties of scattering
amplitudes can be exploited to achieve the decomposition of the
whole amplitude in terms of MI's. In fact, unitarity and factorization,
both at the integral and integrand level, become suitable tools for the
{\it simultaneous} decomposition of group of diagrams
in terms of MI's \cite{Bern:1994cg,Britto:2004nc,Ossola:2006us} 
(see \cite{Britto:2010xq,Ellis:2011cr} for reviews). 
Decomposing the whole amplitude, rather than the
individual diagrams, has the advantage that at each step of the
computation the coefficients of the decomposition reflect the symmetries
of the amplitude. Whereas, the decomposition of each integral
separately may not carry such information, with the drawback of
introducing spurious terms at the intermediate steps of the
calculation, which necessarily disappear from the total result.

The advantage of decomposing amplitudes, rather than individual integrals, in terms of MI's
clearly emerged in the last decade, as it has been the driving principle determining the breakthrough
in the evaluation of next-to-leading order (NLO) virtual corrections
to multi-leg processes (see for instance \cite{Butterworth:2014efa}
and references therein). 
Activities are ongoing to extend the
underlying ideas to higher orders \cite{Mastrolia:2011pr,Kosower:2011ty,Zhang:2012ce,Mastrolia:2012an}.

The evaluation of MI's usually proceeds 
one-by-one in a bottom up approach, starting from the less complicated integrals
and systematically enriching their structure by increasing the number
of internal and external lines.
MI's are functions of the kinematic invariants
built with the external momenta and of
the masses of the involved particles.
Remarkably,  the integration-by-parts (IBP) relations imply that the MI's
obey linear systems of first-order differential equations (DE's) in
the kinematic invariants,
which can be used for the determination of their actual analytic expressions
\cite{Kotikov:1990kg,Remiddi:1997ny,Gehrmann:1999as}.

The simplicity of the whole amplitude with respect to the individual
integrals is reflected also in the fact that the analytic functions present in each MI
do not necessarily appear in the final result, obtained after combining coefficients and MI's altogether. 
Remarkably, Hopf algebra structure (of multiple polylogarithms) can be
used to simplify complicated expressions for multi-loop amplitudes
\cite{Duhr:2012fh}, 
or global symmetries can be used to constrain the minimal set
of variables an amplitude may depend on \cite{Dixon:2014xca}.
In other words, integrals depend on the kinematic invariants enforced by their
external topology and masses, but amplitudes may depend on {\it functions} of
such invariants, which resolve the symmetries that are not manifest at
the diagrammatic level.

In this spirit, the role of differential equations is twofold, because
they are not only a tool for evaluating individual integrals, but also
the {\it location} where investigating how  
the exposed analytic properties of MI's could be transferred to amplitudes.

For any given scattering process the set of MI's is not unique
\cite{Laporta:2001dd}. Rather than considering it a limitation, this
freedom can be exploited in order to select the basis of MI's that can be easier to evaluate.
In the context of the differential equations method 
\cite{Kotikov:1990kg,Remiddi:1997ny,Gehrmann:1999as}, 
reviewed in~\cite{Argeri:2007up,Smirnov:2012gma}, and further
developed in~\cite{Henn:2013pwa,Papadopoulos:2014lla}, 
simplifying the evaluation of MI's means achieving a 
block-triangular form of the system where:
{\it i)} 
the dimensions of the blocks are minimal;
 {\it ii)} 
all analytic properties are exposed;
{\it iii)}  the solution can be determined through an algebraic procedure.

According to the study in Ref.~\cite{Henn:2013pwa}, 
generic systems of DE's for MI's
whose associated matrix depends on rational functions
of the kinematics and $\e$ could be brought in a {\it canonical} form 
where the $\e$-dependence is factorized from the kinematic.
The integration of a canonical system of DE's can be carried out
with algebraic patterns, where the analytic properties of its solution are manifestly
inherited from the matrix associated to the
system, which becomes the kernel of the
representation of the solution in terms of repeated integrations \cite{Goncharov:polylog}.
These novel ideas for evaluating MI's have recently stimulated
several applications~\cite{Henn:2013woa,Henn:2013nsa,Henn:2013wfa,Henn:2013tua,Argeri:2014qva,Caron-Huot:2014lda,Gehrmann:2014bfa,Li:2014bfa,Henn:2014lfa,Caola:2014lpa,Hoschele:2014qsa}.

In Ref.~\cite{Argeri:2014qva}, we proposed to address the 
evaluation of MI's by means of a unitary-like formalism, where Magnus
exponential \cite{Magnus} is responsible for the kinematic {\it evolution} of the
MI's by {\it acting} on the boundary conditions.
In particular, we observed that an initial set of  MI's obeying a systems of
DE's which depends {\it linearly} on $\e$ can be brought in
canonical form after applying a change of basis by means of a
transformation that absorbs the constant term. 
The change from the initial set to the canonical basis of MI's can be implemented with Magnus exponential \cite{Magnus,Blanes:arXiv0810.5488}.
The solution admits an $\e$-expansion in terms of Dyson or equivalently Magnus series,
whose coefficients are written as nested integrations \cite{Goncharov:polylog,Goncharov2001,Remiddi:1999ew,Gehrmann:2001pz,Vollinga:2004sn, Ablinger2011}.

\begin{figure}[t]
\centering
\subfloat[]{\label{Fig:3L1mLadder:a}\includegraphics[width=0.35\textwidth]{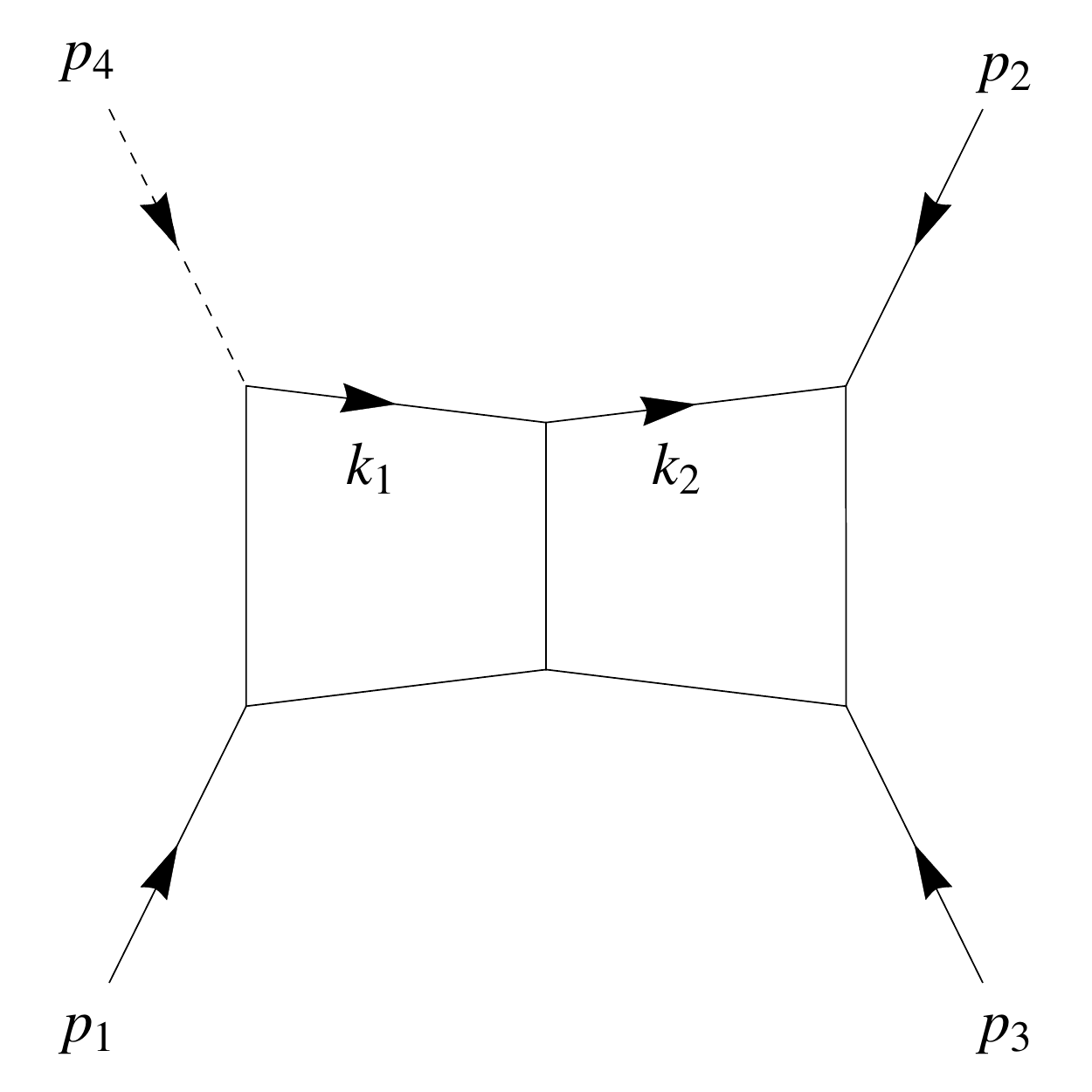}}%
$\quad$%
\subfloat[]{\label{Fig:3L1mLadder:b}\includegraphics[width=0.35\textwidth]{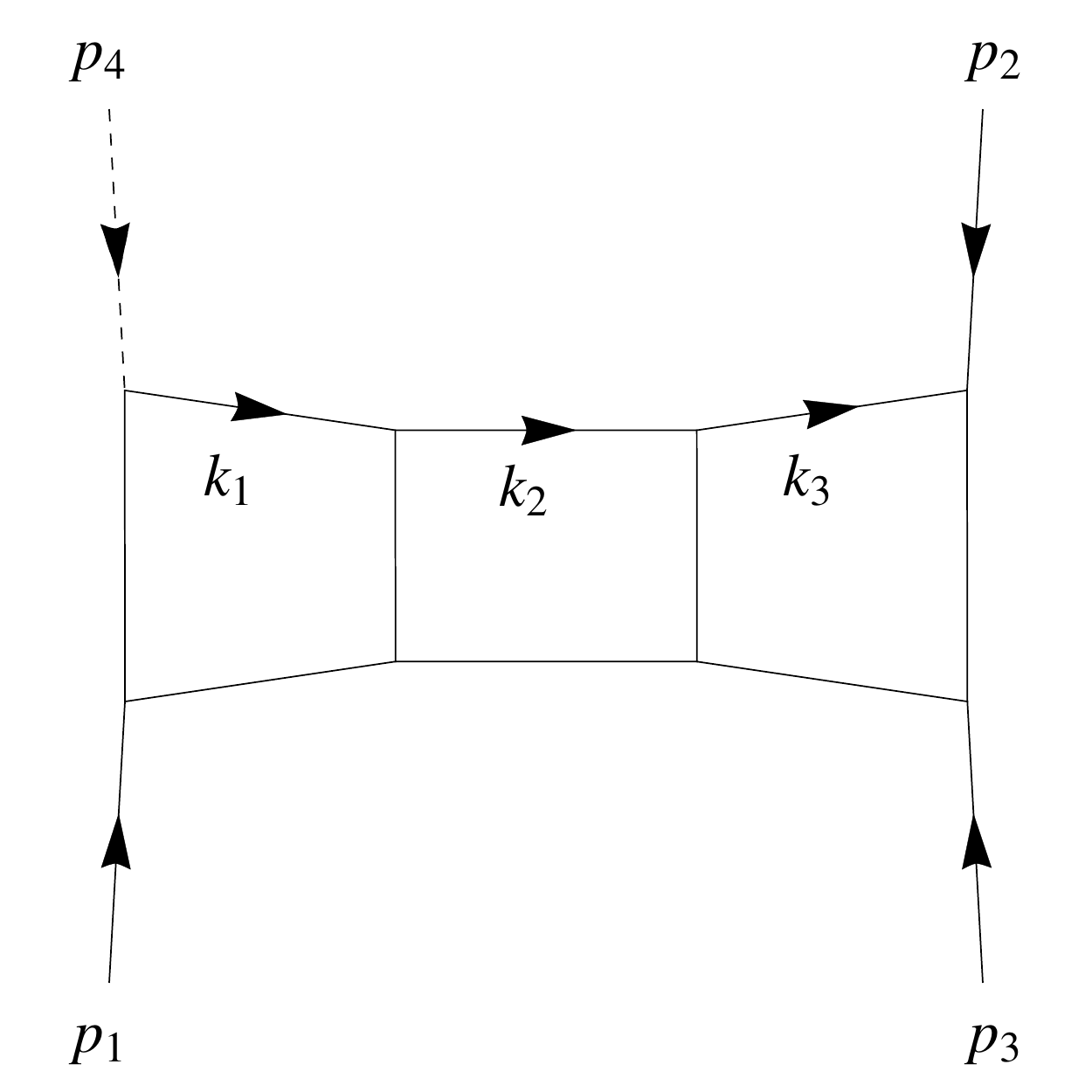}}%
\caption{The two-loop and three-loop
ladder box diagram, with one off-shell leg: 
the solid lines stand for massless particles; the dashed line
represents a massive particle. Momentum conservation is 
$\sum_{i=1}^4p_i = 0$, with $p_i^2 = 0$ $(i=1,\ 2,\ 3)$ and $p_4^2 = m^2$.}
\label{Fig:3L1mLadder}
\end{figure}

In this article, we apply Magnus exponential method to solve the
system of differential equations fulfilled by 85 MI's
required for the determination of the three-loop 
ladder box integrals with one massive leg, shown in
Fig.~\ref{Fig:3L1mLadder:b}, and of the tower of integrals
associated to their subtopologies, depicted in Figs.~\ref{Fig:3loopMIs1}--\ref{Fig:3loopMIs3}.
All propagators are massless.
The solution of the system is finally
determined after fixing the values of the otherwise arbitrary constants
that naturally arise from solving differential equations. In the considered case,
boundary conditions are obtained  by imposing  the {\it regularity} of the MI's  around unphysical
singularities, ruling out the divergent behavior of the general solution of the systems.
 
Among the evaluated 85 MI's, there are 16 vertex-like integrals
with two off-shell legs not yet considered in the literature, 
and 10 one-scale vertex- and bubble-like integrals, which
had already been computed by means of Mellin-Barnes integral
representation \cite{Heinrich:2009be,Heinrich:2007at,Gehrmann:2006wg,Lee:2010cga}.
In general, homogeneous differential equations for single scale
integrals carry only information on the scaling behaviour of the
solution. The determination of the boundary constants
for such differential equations amounts to the
evaluation of the integrals themselves by other means.
Within a multi-scale problem, where integrals may depend on more than
one external invariant, single-scale integrals enter the regularity conditions (or
equivalently could be the limit) of the multi-scale ones~\cite{Argeri:2002wz,Henn:2013nsa}.
These relations, entangling single- and multi-scale functions, can be
exploited to determine the arbitrary constants of the single-scale
integrals, or at least to reduce the number of independent
single-scale integrals that needs to be computed by alternative
methods, other than differential equations.
Therefore, solving multi-scale systems of differential equations
yields the simultaneous determination of single- and multi-scale MI's,
which are finally expressed in terms of a few single-scale MI's, to be independently
provided. In fact, the considered case of 85 MI’s requires only 2 one-scale integrals
as external {\it input}.

The MI's hereby computed can be considered the first contributions to the 
next-to-next-to-next-to-leading order (N$^3$LO) 
virtual corrections to scattering processes like 
the three-jet production from vector boson decay, $V^{*} \to jjj$, 
as well as the Higgs plus one-jet production in gluon fusion, $pp \to Hj$,  
currently computed at NNLO accuracy in Refs.~\cite{Garland:2001tf,Garland:2002ak,GehrmannDeRidder:2009dp,Weinzierl:2009ms}
and Refs.~\cite{Gehrmann:2011aa,Boughezal:2013uia}
respectively. 
The collinear limits of the MI's we present enter the computation of the three-loop
one-particle splitting amplitudes, currently known at two-loop order~\cite{Badger:2004uk}.
Such amplitudes also serve as an ingredient for the derivation,
following the procedure of Ref.~\cite{Kosower:2003np},
of the N$^3$LO Altarelli-Parisi splitting kernel.

The results of the considered box-type integrals and of the
tower of vertex- and bubble-type integrals associated to
subtopologies are given as a Taylor series expansion in $\epsilon$. 
The coefficients of the series are expressed in terms of uniform weight
combinations of transcendental constants and generalised harmonic polylogarithms 
\cite{Remiddi:1999ew,Gehrmann:2001pz,Gehrmann:2001jv}
up to weight six.

We also present the calculation of the two-loop one-mass
planar box diagram in Fig.~\ref{Fig:3L1mLadder:a},
giving the result for the corresponding MI's, whose topologies are
depicted in Fig.~\ref{Fig:3loopMIs1}. 
We provide higher orders $\epsilon$-expansions w.r.t. the expressions available in the literature~\cite{Gehrmann:1999as,Gehrmann:2000zt}.

We used the computer code \texttt{Reduze 2} \cite{Studerus:2009ye,vonManteuffel:2012np}
for solving the system of integration-by-parts relations
and generating the system of differential equations. The analytic results of the
MI's at two- and three-loop can be numerically
evaluated by means of \texttt{GiNaC} \cite{Bauer:2000cp,Vollinga:2004sn} and are
found in agreement with the outcome of the direct numerical
integrations carried out by \texttt{FIESTA 3}
\cite{Smirnov:2008py,Smirnov:2013eza} (within per-mille level of accuracy).

The paper is organized as follows. 
In Section~\ref{sec:diffeq}, we discuss the generic form of the systems of
differential equations which are required to compute the ladder box integrals
with one massive-leg at two- and three-loop. There we show
how the canonical systems are constructed by using Magnus exponential matrices.
In Sections~\ref{sec:2loop} and \ref{sec:3loop}, we present the
canonical bases for the two- and three-loop cases respectively.
In Section~\ref{sec:boundaries}, the limiting conditions for fixing the boundary are
explained, showing how meaningful relations for single-scale integrals emerge.

In the Appendices~\ref{App:2loopMatrices} and~\ref{App:3loopMatrices}, we collect the representative matrices of the
canonical systems.
Appendix~\ref{App:3loopg83} contains the analytic expression of the three-loop one-mass
scalar-box integral.

The sets of 18 two-loop MI's and of 85 three-loop MI's
are collected in two ancillary files, {\tt <2loopMIs.m>} and 
{\tt <3loopMIs.m>} respectively.

%% file: diffeq.tex
\section{Differential equations and Magnus Exponential}
\label{sec:diffeq}

All Feynman integrals belonging to the topologies which we consider in the present work can be
defined in terms of the following sets of denominators $\mathcal{D}_n$ ($k_i$ are the loop momenta).
For the two-loop ladder diagram in Fig.~\ref{Fig:3L1mLadder:a}, one has
\begin{gather}
  \mathcal{D}_1 = k_1^2, \quad
  \mathcal{D}_2 = k_2^2, \quad
  \mathcal{D}_3 = (k_1-k_2)^2, \quad
  \mathcal{D}_4 = (k_2+p_2)^2, \nonumber \\
  \mathcal{D}_5 = (k_1+p_2+p_3)^2, \quad
  \mathcal{D}_6 = (k_2+p_2+p_3)^2, \quad
  \mathcal{D}_7 = (k_1+p_1+p_2+p_3)^2,\\
  \intertext{supplemented by the auxiliary denominators}
  \mathcal{D}_8 = (k_1+p_2)^2, \quad
  \mathcal{D}_9 = (k_2+p_1+p_2+p_3)^2.
\end{gather}
For the three-loop ladder diagram in Fig.~\ref{Fig:3L1mLadder:b} one has instead
\begin{gather}
  \mathcal{D}_1 = k_1^2,\quad
  \mathcal{D}_2 = k_2^2,\quad
  \mathcal{D}_3 = k_3^2,\quad
  \mathcal{D}_4 = (k_1-k_2)^2, \nonumber\\
  \mathcal{D}_5 = (k_2-k_3)^2,\quad
  \mathcal{D}_6 = (k_3+p_2)^2,\quad
  \mathcal{D}_7 = (k_1+p_2+p_3)^2,\nonumber\\
  \mathcal{D}_8 = (k_2+p_2+p_3)^2,\quad
  \mathcal{D}_9 = (k_3+p_2+p_3)^2,\quad
  \mathcal{D}_{10} = (k_1+p_1+p_2+p_3)^2,\\
  \intertext{supplemented by the auxiliary denominators}
  \mathcal{D}_{11} = (k_1+p_2)^2,\quad
  \mathcal{D}_{12} = (k_2+p_2)^2,\quad
  \mathcal{D}_{13} = (k_2+p_1+p_2+p_3)^2,\nonumber\\
  \mathcal{D}_{14} = (k_3+p_1+p_2+p_3)^2,\quad
  \mathcal{D}_{15} = (k_3-k_1)^2.
\end{gather}
In the following we consider $\ell$-loop Feynman integrals built out of $p$ of the above denominators, each raised to some integer power, of the form
\begin{gather}
  \int \widetilde{\mathrm{d}^d k_1} \ldots \widetilde{\mathrm{d}^d k_\ell} \, \frac{1}{\mathcal{D}_{a_1}^{n_1} \ldots \mathcal{D}_{a_p}^{n_p}},
\end{gather}
where the integration measure is defined as 
\begin{gather}
  \widetilde{\mathrm{d}^dk_i} \equiv \frac{\mathrm{d}^d k_i}{(2\pi)^d} \left(\frac{i \, S_\eps}{16 \pi^2} \right)^{-1} (-m^2)^{\eps},
  \shortintertext{with}
  S_\eps \equiv (4\pi)^\eps \, \frac{\Gamma(1+\eps)\, \Gamma^2(1-\eps)}{\Gamma(1-2\eps)}.
  \label{eq:intmeasure}
\end{gather}
The two- and three-loop MI's,
respectively depicted in Fig.~\ref{Fig:2loopMIs} and Figs.~\ref{Fig:3loopMIs1}--\ref{Fig:3loopMIs3},
are functions of the kinematic variables
\begin{gather}
  s=(p_2+p_3)^2, \quad t=(p_1+p_3)^2,  \quad u=(p_1+p_2)^2,  \quad m^2=(p_1+p_2+p_3)^2,
\end{gather}
with
$p_i^2=0 \ (i=1,2,3)$ and $p_4^2=(p_1+p_2+p_3)^2$, fulfilling $s+t+u=m^2$.
For convenience, we define the dimensionless ratios
\begin{gather}
  x \equiv \frac{s}{m^2}, \quad
  y \equiv \frac{t}{m^2}, \quad
  z \equiv \frac{u}{m^2}, \quad \text{ with }\;x+y+z = 1.
\end{gather}
For planar topologies, like the ones we consider, one can always choose the Euclidean kinematic region
$m^2,\,s,\,t,\,u <0$ such that the MI's are real. For definiteness, we work in the region $0<y<1$, $0<x<1-y$
(or equivalently $0<x<1$, $0<y<1-x$). The analytic continuation of our results to regions of physical interests
can be performed by generalizing to higher weights the procedure of Ref.~\cite{Gehrmann:2002zr}.

The sets of MI's we choose to work with (see Sections~\ref{sec:2loop} and~\ref{sec:3loop}) obey
$\e$-linear systems of first order differential equations in the kinematic variables ($\sigma=x,y$),
\begin{align}
  \der{\sigma} f(\e,m^2,x, y) &= A_{\sigma}(\e,m^2,x,y)\,
  f(\e,m^2,x,y) ,
  \label{eq:MIdiffeq}
  \shortintertext{with}
  A_{\sigma}(\e,m^2,x,y) &=
  A_{\sigma,0}(m^2,x,y)+ \e\, A_{\sigma,1}(m^2,x,y) .
\end{align}
Given that $m^2$ is the only dimensionful variable, the differential equation in $m^2$ is related to the scaling
equation. A factor $(-m^2)^{-\ell \eps }$ is already included in integration measure~\eqref{eq:intmeasure}, therefore
for each $f_i$ we simply have
\begin{align}
  \der{m^2} f_i(m^2,x, y,\e) = -\frac{n_i}{m^2} f_i(\e,m^2,x,y) ,
  \label{eq:scaling}
\end{align}
where $n_i$ is the dimension of the integral $f_i$ in units of a squared mass.

\subsection{Magnus exponential}

We can use Magnus exponential \cite{Magnus,Argeri:2014qva} to define a matrix $B$ that
implements a change of basis $f \to g$,
\begin{gather}
  f \equiv B g, \qquad
  {\hat A}_\sigma \equiv  B^{-1} A_\sigma B - B^{-1} \partial_\sigma B,
  \label{eq:canonicaltfm}
\end{gather}
such that the new basis of MI's fulfills
a canonical system of differential equations,
\begin{gather}
  \der{\sigma} g(\e,x,y) = \e\, {\hat A}_{\sigma}(x,y)\, g(\e,x,y),
  \label{eq:cansys1}
\end{gather}
where the dependence on $\e$ is factorized from that on the kinematic variables~\cite{Henn:2013pwa}.
Accordingly, $g$ is called {\it canonical basis} of MI's.

Following Ref.~\cite{Argeri:2014qva},
in order to build the matrix $B$ of~\eqref{eq:canonicaltfm},
let us introduce Magnus exponential matrix
\cite{Magnus,Blanes:arXiv0810.5488}. For a
generic matrix ${M}=M(t)$, this is defined as
\begin{gather}
  e^{\Omega[M(t)]} ,
\end{gather}
being the solution of the following matrix differential equation,
\begin{gather}
  \der{t} e^{\Omega[M(t)]} = M(t) \, e^{\Omega[M(t)]} .
  \label{eq:MagnusDiffEq}
\end{gather}
The exponent $\Omega[M(t)]$ is given as a series of matrices obtained by repeated
integrations, using $M(t)$ as a kernel,
\begin{gather}
  \Omega[ M(t) ] \equiv \sum_{n=1}^\infty \Omega_n[ M(t) ] ,
  \label{eq:Magnus}
\end{gather}
where the first three terms of the series expansion read,
\begin{align}
  \Omega_1[M(t)] &= \int_{t_0}^t d\tau_1 \, M(\tau_1),\nonumber  \\
  \Omega_2[M(t)] &= \frac{1}{2} \int_{t_0}^t \! d\tau_1
    \int_{t_0}^{\tau_1}\! d\tau_2 \, \bigl[M(\tau_1), M(\tau_2)\bigr], \label{eq:Magnus:exp}\\
  \Omega_3[M(t)] &= \frac{1}{6} \int_{t_0}^t \! d\tau_1
    \int_{t_0}^{\tau_1}\! d\tau_2 \int_{t_0}^{\tau_2} \! d\tau_3  \,
    \bigl[M(\tau_1),\bigl[M(\tau_2),M(\tau_3)\bigr]\bigr] +
    \bigl[M(\tau_3),\bigl[M(\tau_2),M(\tau_1)\bigr]\bigr] . \nonumber
\end{align}

\subsection{Canonical transformation}

The basis change $B$, which brings the systems in the canonical form~\eqref{eq:cansys1}, can in general be
obtained in both the two- and three-loop cases as the product of five exponential matrices, each
implementing a change of basis according to~\eqref{eq:canonicaltfm},
applied in successions. The dependence of each matrix on
$x$ and~$y$ is understood.
\begin{enumerate}

\item 
First of all we rescale all the MI's by appropriate factors of $m^2$ in order to work with dimensionless integrals.
This can be conveniently implemented by a Magnus exponential of the diagonal matrix $A_{m^2\!,0}$
\begin{gather}
A_\sigma \to A^{[0]}_{\sigma}, \qquad B^{[0]} \equiv e^{\magnus{A_{m^2\!,0}}} .
\end{gather}
As a result of this transformation, $A^{[0]}_{m^2}$ vanishes and the MI's do not depend anymore on $m^2$.

\item 
Then we decompose $A^{[0]}_{x,0}$ in a diagonal term $D^{[0]}_{x,0}$ and
a off-diagonal one $N^{[0]}_{x,0}\,$,
\begin{gather}
  A^{[0]}_{x,0} = D^{[0]}_{x,0} + N^{[0]}_{x,0}\, ,
\end{gather}
and use only the diagonal part for the next basis change,
\begin{gather}
  A^{[0]}_{\sigma} \to  A^{[1]}_{\sigma} , \qquad B^{[1]} \equiv
  e^{\magnus{D^{[0]}_{x,0}}} .
\end{gather}

\item 
We repeat as before and split $A^{[1]}_{y,0}$ into its
diagonal and off-diagonal parts,
\begin{gather}
  A^{[1]}_{y,0} = D^{[1]}_{y,0} + N^{[1]}_{y,0} \, ,
\end{gather}
and we build the Magnus exponential again using the diagonal part
\begin{gather}
  A^{[1]}_{\sigma} \to  A^{[2]}_{\sigma} , \qquad B^{[2]} \equiv
  e^{\magnus{D^{[0]}_{y,0}}} .
\end{gather}

\item 
Now the matrix $A^{[2]}_{x,0}\,$, has no diagonal term left,
\begin{gather}
  A^{[2]}_{x,0} =  N^{[2]}_{x,0} \, ,
\end{gather}
therefore we build the Magnus exponential of the off-diagonal part
\begin{gather}
  A^{[2]}_\sigma \to A^{[3]}_{\sigma}  , \qquad B^{[3]} \equiv
  e^{\magnus{N^{[2]}_{x,0}}} .
\end{gather}

\item 
The matrix $A^{[3]}_{y,0}$ has no diagonal term as well,
\begin{gather}
  A^{[3]}_{y,0} =  N^{[3]}_{y,0} \, ,
\end{gather}
so we can define the last basis change
\begin{gather}
  A^{[3]}_\sigma \to A^{[4]}_{\sigma} , \qquad B^{[4]} \equiv
  e^{\magnus{N^{[3]}_{y,0}}} .
\end{gather}
\end{enumerate}
After the last transformation we observe that
\begin{gather}
  A^{[4]}_{x,0} = 0 = A^{[4]}_{y,0} \, .
\end{gather}
This means that the basis change of~\eqref{eq:canonicaltfm}, with the matrix $B$ given by
\begin{eqnarray}
  B & \equiv & B^{[0]} B^{[1]} B^{[2]} B^{[3]} B^{[4]} = 
  e^{\magnus{A_{m^2\!,0}}} \,
  e^{\magnus{D^{[0]}_{x,0}}} \,
  e^{\magnus{D^{[1]}_{y,0}}} \,
  e^{\magnus{N^{[2]}_{x,0}}} \,
  e^{\magnus{N^{[3]}_{y,0}}}, 
\label{eq:genericBmatrix}
\end{eqnarray}
\emph{absorbs} the constant terms of ${A}_{x}$ and ${A}_{y}$ in
the $\e$-linear systems in~\eqref{eq:MIdiffeq} and brings them to the canonical form~\eqref{eq:cansys1}:
\begin{align}
  A_\sigma(\eps,m^2,x,y) & \to \eps \hat{A}_{\sigma}(x,y).
 \label{eq:canonicalxy}
\end{align}

\subsection{Canonical system}

The two partial-derivative systems satisfied by the new set of MI's, $g = B^{-1}f$, can be conveniently
combined in an exact differential form,
\begin{gather}
  d g(\e,x,y) = \e \, d{\hat {\cal A}}(x,y) \, g(\e,x,y) , \qquad
  d{\hat {\cal A}} \equiv  {\hat A}_{x} dx +  {\hat A}_{y} dy ,
  \label{eq:cansys2}
\intertext{where $\hat {\cal A}(x,y)$ is logarithmic in the variables $x$ and  $y$,}
  \hat {\cal A}(x,y) =
  M_1 \log(x) + M_2 \log(1-x)+M_3 \log(y)+M_4 \log(1-y) +{} \nonumber \\
  {}+M_5 \log\bigg(\frac{x+y}{x}\bigg)
  +M_6  \log\bigg(\frac{1-x-y}{1-x}\bigg).
  \label{eq:GenCanSys}
\end{gather}
The matrices $M_i$ are $n \times n$ sparse matrices with purely
rational entries, where $n$ is the number of MI's. The value of $n$ depends on
the number of loops, and it amounts to $n=4,\,18,\,85$, respectively at
one-, two- and three-loop.
The arguments of the logarithms are defined as {\it letters}, and
the set of letters
\begin{gather}
  \{ x,\, 1-x,\, y,\, 1-y,\, x+y, \, 1-x-y\}
  \label{def:alphabet}
\end{gather}
constitutes the {\it alphabet}.
We observe that this alphabet is common both to the two- and the three
loop cases.
At one-loop \cite{Gehrmann:1999as}, although not shown here, $M_5$ is absent,
while at two-loop $M_5$ has only one non-vanishing entry.

The solution of~(\ref{eq:cansys1}, \ref{eq:cansys2}) can be
expressed as a Dyson series in
$\epsilon$,
\begin{align}
  g(x,y,\epsilon) &= \biggl(1 + \sum_{n=1}^\infty \e^n D^{(n)}(x,y) \biggr)\,
  g_0(\epsilon) ,
  \label{eq:sol:gen:Dyson}
\intertext{where the (matrix) coefficients of the series can be written as the iterated line integral,}
   D^{(n)}(x,y) &\equiv
  \int_{\gamma}
  d{\hat {\cal A}_1} \,
  d{\hat {\cal A}_2}
  \cdots
  d{\hat {\cal A}_n}\,,
  \label{eq:sol:Dyson}
\intertext{where $d{\hat {\cal A}_i}\equiv d{\hat {\cal A}(x_i,y_i)}$.
Equivalently, the solution admits a representation in terms of the Magnus
exponential}
  g(x,y,\epsilon) &= e^{\Omega[\e\, d{\hat {\cal A}}](x,y)}\,  g_0(\epsilon) ,
  \label{eq:sol:Magnus}
\end{align}
where the vector $g_0(\e) \equiv g(x_0,y_0,\e)$ corresponds to the boundary values of the MI's.
This form is very suggestive, as Magnus exponential can be considered as an
{\it evolution} operator, like in the unitary formalism, that brings the
MI's $g$ from their initial, boundary values to the considered point
in the $(x,y)$-plane\footnote{In this case, the evolution has to be understood
  like the variation w.r.t.\ the
  kinematic invariants that are the variables of the
  system of differential equations obeyed by MI's, rather than 
  the quantum-mechanical time-evolution. Moreover, the matrices
  representing the systems of differential equations for MI's are not unitary.}.

For definiteness, we integrate the exact differential form~\eqref{eq:cansys2}
from an arbitrary point $(x_0,y_0)$ to $(x,y)$, along the broken path composed of the two
segments in which one of the variables is kept constant.
The integration is performed order by order in $\eps$, up to a multiplicative
vector of unknown constants. The latter are fixed by requiring the regularity of $g(x,y,\eps)$ at the
pseudothresholds (see Sec.~\ref{sec:boundaries}).

The solution of the canonical system in~\eqref{eq:GenCanSys} with
the coefficient matrix given by~\eqref{eq:cansys2}
can be naturally expressed in terms of Goncharov's multiple polylogarithms ($G$-polylogarithms, for short)
\cite{Goncharov:polylog,Remiddi:1999ew,Gehrmann:2001pz,Vollinga:2004sn},
\begin{align}
  G({\vec w}_{n} ; x) &\equiv G(w_1, {\vec w}_{n-1} ; x) \equiv
  \int_0^x dt \frac{1}{t-w_1}
  G({\vec w}_{n-1};t) , \\
  G(\vec{0}_n;x)& \equiv \frac{1}{n!}\text{log}^n(x) ,
\end{align}
with ${\vec w}_n$ being a vector of $n$ arguments. The number $n$ is referred to as
the {\it weight} of $G({\vec w}_{n} ; x)$ and amounts to the number of iterated integrations
needed to define it.
Equivalently one has
\begin{equation}
  \der{x} G(\vec{w}_{n}; x) = \der{x} G(w_1,\vec{w}_{n-1}; x) = \frac{1}{x-w_1} G(\vec{w}_{n-1};x).
\end{equation}
$G$-polylogarithms fulfill shuffle algebra relations of the type
\begin{equation}
  G(\vec{m};x)\,G(\vec{n};x) = G(\vec{m};x) \shuffle G(\vec{n};x)
  = \:\sum_{\mathclap{\vec{p}=\vec{m} \shuffle \vec{n}}}\: G(\vec{p};x),
\end{equation}
where shuffle product $\vec{m} \shuffle \vec{n}$ denotes all possible merges
of $\vec{m}$ and $\vec{n}$ preserving their respective orderings.
Because of shuffle relations, for a given alphabet and a given
weight one can identify a minimal basis of $G$-polylogarithms.
For the calculation of the considered two- and three-loop MI's
the alphabet in~\eqref{def:alphabet} corresponds to the
set of weights $W=\{0,\,1,\,-x,\,1-x \}$.
The number of basis elements depending on the weight $n$ and the alphabet size $\alpha$
is given by the \emph{Witt formula}:
\begin{gather}
  N(n,\alpha) = \frac{1}{n} \sum_{d|n} {\mu}(d)\, \alpha^{\frac{n}{d}},
\end{gather}
where $\mu$ denotes the M\"obius function and the sum is done over all divisors of $n$.
For reference we give values of $N(n,\alpha)$ relevant for this calculation in Tab.~\ref{tab:2dhpl}.

\begin{table}[t]
\begin{center}
\begin{tabular}{ c | r r r }
\hline
\hline
{ weight} & { $\alpha=2$} & { $\alpha=3$} & { $\alpha=4$}  \\
\hline
1 & 2 & 3 & 4 \\
2 & 1 & 3 & 6 \\
3 & 2 & 8 & 20 \\
4 & 3 & 18 & 60 \\
5 & 6 & 48 & 204 \\
6 & 9 & 116 & 670 \\
\hline
\hline
\end{tabular}
\end{center}
\caption{Size of a basis of $G$-polylogarithms.}
\label{tab:2dhpl}
\end{table}

%% file: 2loop.tex
\begin{figure}[t]
\begin{center}
\includegraphics[width=0.9\textwidth]{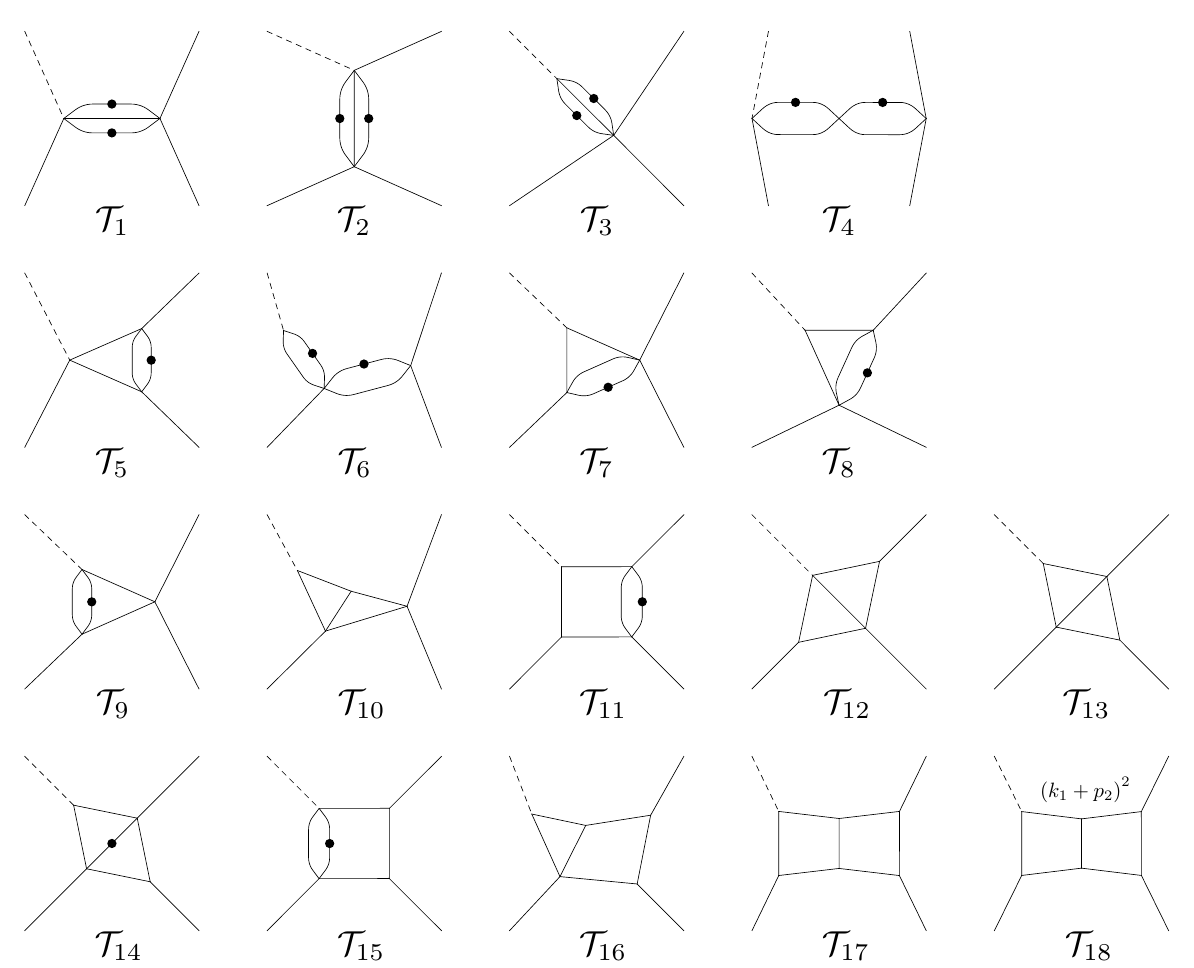}
\end{center}
\caption{Two-loop Master Integrals $\{ {\cal T}_i \}_{i=1,\ldots,18}$.
The solid lines stand for massless particles; the dashed line represents
a massive particle; dots indicate squared propagators;
  numerators may appear as indicated ($p_{ij}\equiv p_i+p_j$).}
\label{Fig:2loopMIs}
\end{figure}

\section{Two-Loop Master Integrals}
\label{sec:2loop}

A set of 18 two-loop MI's for the planar massless box integrals with
one massive leg of Fig.~\ref{Fig:3L1mLadder:a} was first computed in Refs.~\cite{Gehrmann:1999as,Gehrmann:2000zt}. 
We present the calculation of an alternative, yet compatible set of MI's.
We begin by choosing the following basis
\begin{align*}
f_1&=\eps^2 \, \top{1}  &
f_2&=\eps^2 \, \top{2}  &
f_3&=\eps^2 \, \top{3}  \\
f_4&=\eps^2 \, \top{4} &
f_5&=\eps^3 \, \top{5}  &
f_6&=\eps^2 \, \top{6}  \\
f_7&=\eps^3 \, \top{7}   &
f_8&=\eps^3 \, \top{8}  &
f_9&=\eps^3 \, \top{9}  \\
f_{10}&=\eps^4 \, \top{10}  &
f_{11}&=\eps^3 \, \top{11}  &
f_{12}&=\eps^4 \, \top{12}   \\
f_{13}&=\eps^4 \, \top{13}  &
f_{14}&=\eps^3 \, \top{14}  &
f_{15}&=\eps^3 \, \top{15}  \\
f_{16}&=\eps^4 \, \top{16}  &
f_{17}&=\eps^4 \, \top{17}  &
f_{18}&=\eps^4 \, \top{18}  \stepcounter{equation}\tag{\theequation}\label{def:2loopBasisMIs}
\end{align*}
where the integrals $\top{i}$ are depicted in Fig.~\ref{Fig:2loopMIs}.
The set $\{ f \}_{i=1,\ldots,18}$ is chosen to obey a $\e$-linear system
  of differential equations in $x$ and $y$, which, as previously described, can be cast in canonical form by
  means of Magnus exponentials. 
In this case, the canonical transformation $B$, generically defined in~\eqref{eq:genericBmatrix}, reduces to 
$
B \equiv e^{\magnus{A_{m^2\!,0}}} \, e^{\magnus{D^{[0]}_{x,0}}} \, e^{\magnus{D^{[1]}_{y,0}}} ,
$
because after the first three transformations $N^{[2]}_{x,0} =
N^{[2]}_{y,0} = 0$.
The canonical basis $\{ g \}_{i=1,\ldots,18}$ reads,
\begin{align*}
g_1   &= s  \,  f_1 &
g_2   &= t  \,  f_2  &
g_3   &= m^2 \, f_3  \\
g_4   &= s^2 \, f_4 &
g_5   &= s  \,  f_5  &
g_6   &= m^2 \, s \,     f_6  \\
g_7   &= \lambda_s \, f_7  &
g_8   &= \lambda_t \, f_8  &
g_9   &= \lambda_s \, f_9  \\
g_{10}&= \lambda_s \, f_{10}  &
g_{11}&= s \, t \, f_{11}  &
g_{12}&= u  \,    f_{12}  \\
g_{13}&= \lambda_u \, f_{13}  &
g_{14}&= s \, t \, f_{14}  &
g_{15}&= s \, t \, f_{15}  \\
g_{16}&= s \, \lambda_t \, f_{16}  &
g_{17}&= s^2 \, t \, f_{17}  &
g_{18}&= s \, \lambda_s \, f_{18}  \stepcounter{equation}\tag{\theequation}\label{def:2loopCanonicalMIs}
\end{align*}
where
$\lambda_a = \left(m^2-a\right)$.
The sparse matrices $M_i$ $(i=1,\ldots,6)$ appearing in the corresponding canonical
system in~\eqref{eq:cansys2} and~\eqref{eq:GenCanSys} are given in Appendix~\ref{App:2loopMatrices}.

%% file: 3loop.tex
\begin{figure}[t]
\begin{center}
\includegraphics[width=0.9\textwidth]{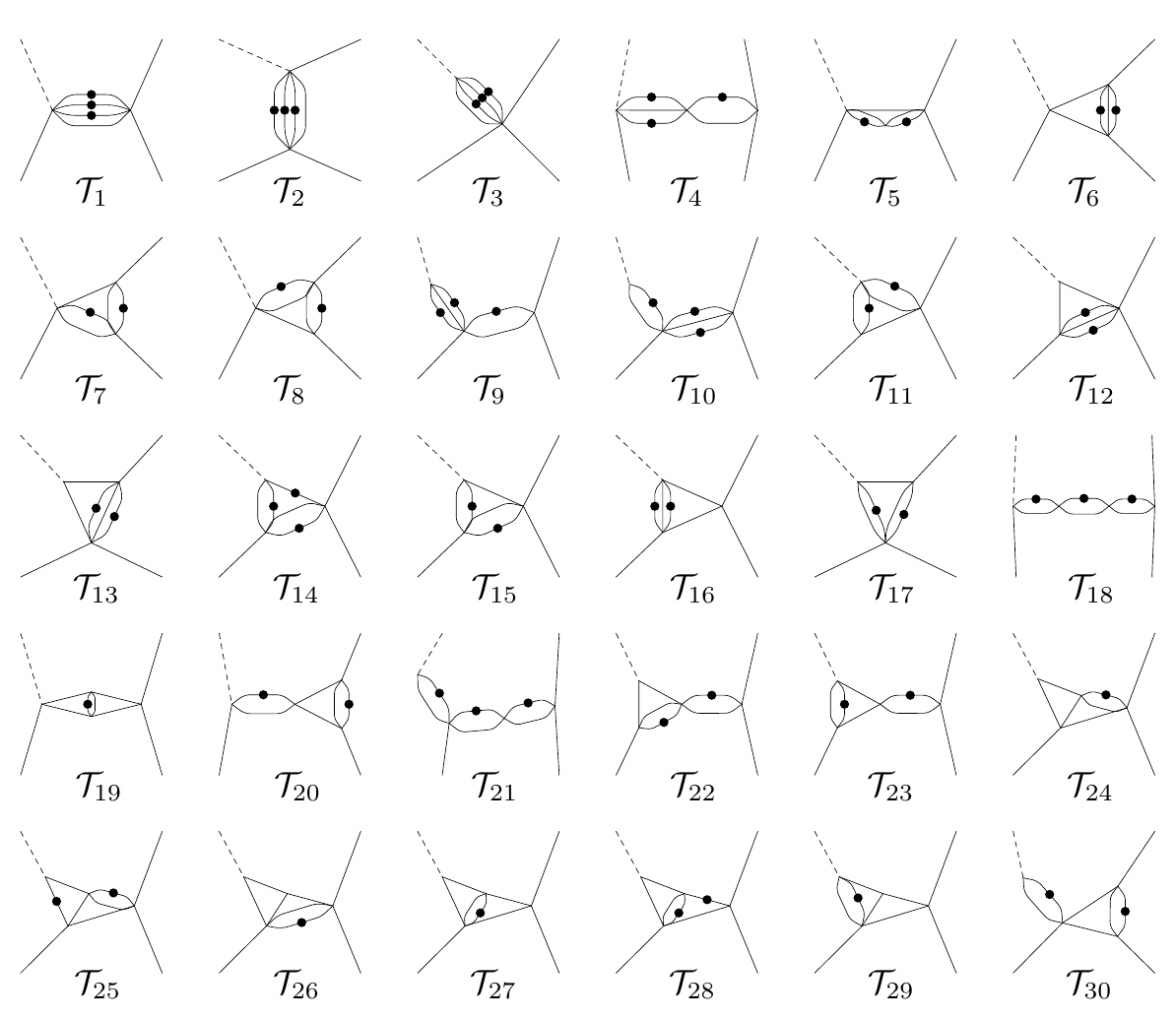}
\end{center}
\caption{Three-loop Master Integrals $\{ {\cal T}_i \}_{i=1,\ldots,30}$.
The solid lines stand for massless particles; the dashed line represents
a massive particle; dots indicate squared propagators;
  numerators may appear as indicated ($p_{ij}\equiv p_i+p_j$).
See also Figs. \ref{Fig:3loopMIs2} and \ref{Fig:3loopMIs3}.}
\label{Fig:3loopMIs1}
\end{figure}

\begin{figure}
\begin{center}
\includegraphics[width=0.9\textwidth]{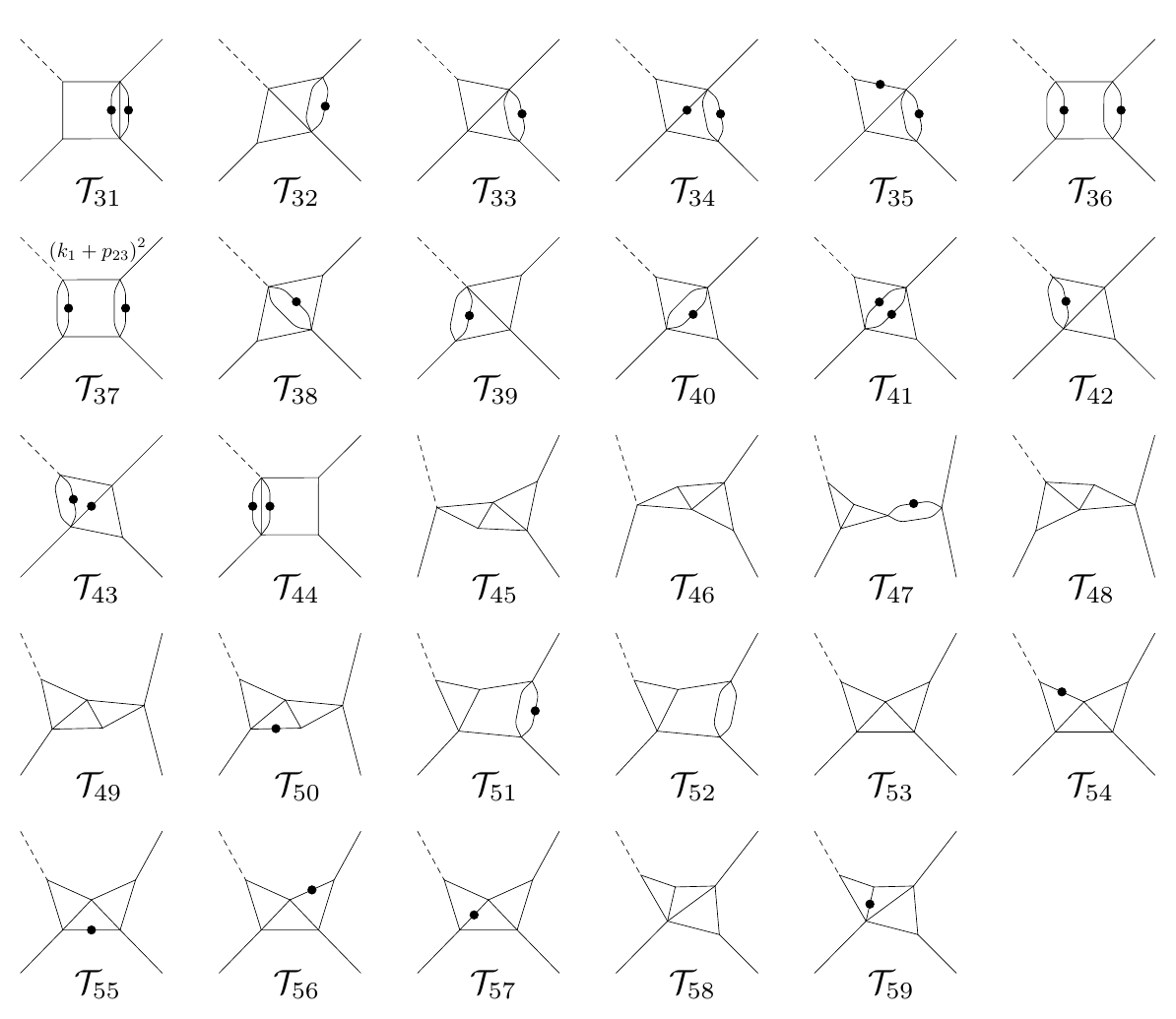}
\end{center}
\caption{Three-loop Master Integrals $\{ {\cal T}_i \}_{i=31,\ldots,59}$.
The solid lines stand for massless particles; the dashed line represents
a massive particle; dots indicate squared propagators;
  numerators may appear as indicated ($p_{ij}\equiv p_i+p_j$). 
See also Figs. \ref{Fig:3loopMIs1} and \ref{Fig:3loopMIs3}.}
\label{Fig:3loopMIs2}
\end{figure}

\begin{figure}
\begin{center}
\includegraphics[width=0.9\textwidth]{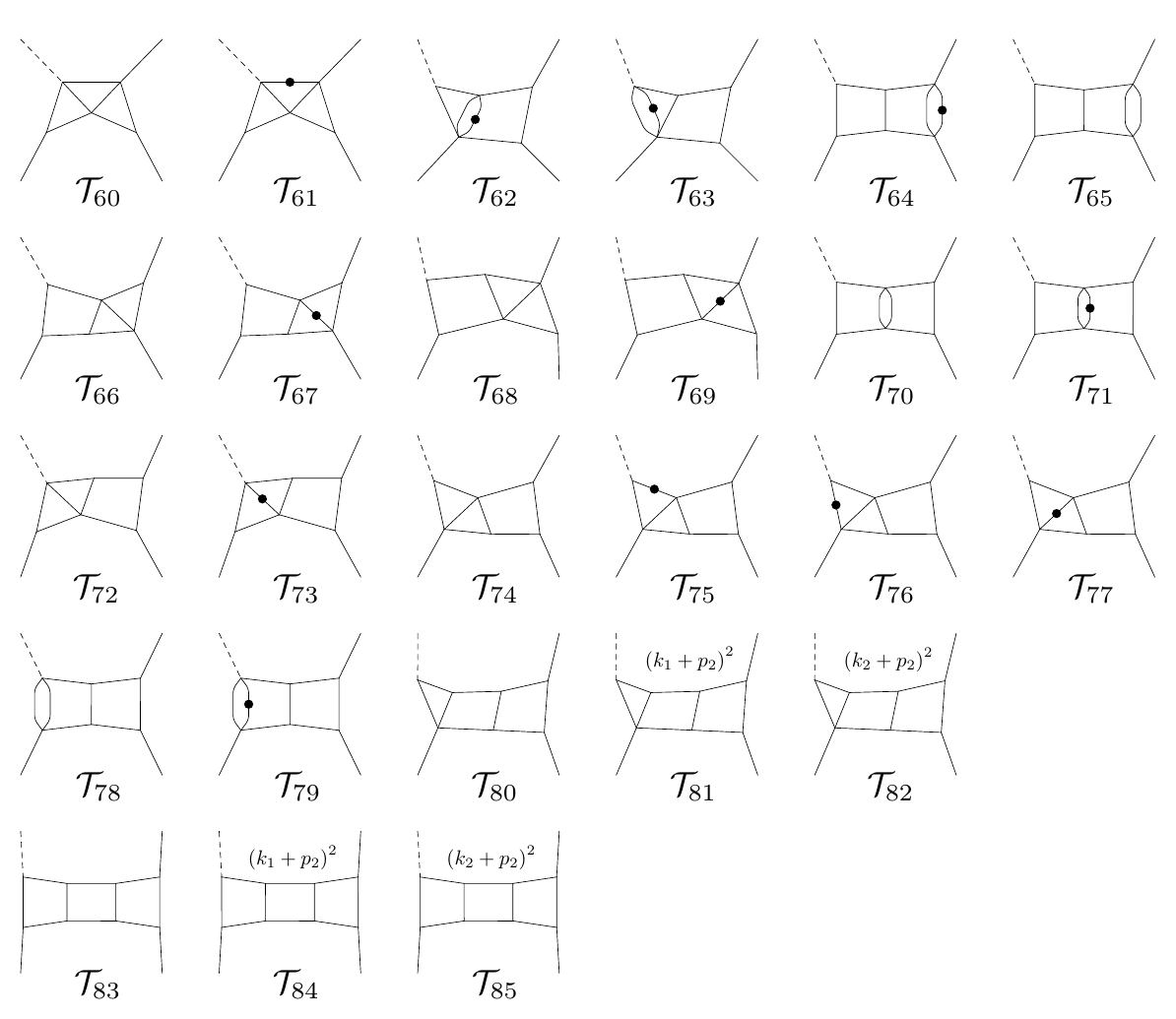}
\end{center}
\caption{Three-loop Master Integrals $\{ {\cal T}_i \}_{i=60,\ldots,85}$.
The solid lines stand for massless particles; the dashed line represents
a massive particle; dots indicate squared propagators;
  numerators may appear as indicated ($p_{ij}\equiv p_i+p_j$). 
See also Figs. \ref{Fig:3loopMIs1} and \ref{Fig:3loopMIs2}.}
\label{Fig:3loopMIs3}
\end{figure}

\section{Three-Loop Master Integrals }
\label{sec:3loop}

The calculation of the three-loop ladder box integrals with one
massive leg, represented in Fig.~\ref{Fig:3L1mLadder:b}, is the main, new result of this paper. 
After performing the automatic reduction by means of the computer code
{\sc Reduze2} \cite{Studerus:2009ye,vonManteuffel:2012np}, we identify
a set of 85 MI's, 
{\allowdisplaybreaks[1]
\begin{align*}
f_1 &= \epsilon ^3  \, \top{1}   &
f_2&=\epsilon ^3 \, \top{2}  &
f_3&=\epsilon ^3 \, \top{3}  \\
f_4&=\epsilon ^3 \, \top{4}  &
f_5&=\epsilon ^3 \, (1+2 \epsilon ) \, \top{5}  &
f_6&=\epsilon^4 \, \top{6}  \\
f_7&=\epsilon ^4 \, \top{7}  &
f_8&=\epsilon ^4 \, \top{8}  &
f_9&=\epsilon ^3 \, \top{9} \\
f_{10}&=\epsilon ^3 \, \top{10}  &
f_{11}&=\epsilon ^4 \, \top{11}  &
f_{12}&=\epsilon^4 \, \top{12} \\
f_{13}&=\epsilon ^4 \, \top{13}  &
f_{14}&=\epsilon ^3 \, \top{14}  &
f_{15}&=\epsilon ^4 \, \top{15}  \\
f_{16}&=\epsilon ^4 \, \top{16}   &
f_{17}&=\epsilon ^4 \, \top{17} &
f_{18}&=\epsilon ^3 \, \top{18}  \\
f_{19}&=\epsilon ^4 \, (1-2 \epsilon )  \, \top{19}   &
f_{20}&=\epsilon ^4 \, \top{20}   &
f_{21}&=\epsilon ^3 \, \top{21}   \\
f_{22}&=\epsilon^4 \, \top{22}  &
f_{23}&=\epsilon ^4 \, \top{23}    &
f_{24}&=\epsilon ^5 \, \top{24}  \\
f_{25}&=\epsilon ^4 \, \top{25}   &
f_{26}&=\epsilon ^5 \, \top{26}   &
f_{27}&=\epsilon ^5 \, \top{27}  \\
f_{28}&=\epsilon ^3 \, (1+2 \epsilon) \, \top{28}     &
f_{29}&=\epsilon ^5 \, \top{29}   &
f_{30}&=\epsilon ^4 \, \top{30}  \\
f_{31}&=\epsilon ^4 \, \top{31}    &
f_{32}&=\epsilon^5 \, \top{32}    &
f_{33}&=\epsilon ^5 \, \top{33}  \\
f_{34}&=\epsilon ^4 \, \top{34}   &
f_{35}&=\epsilon ^4 \, \top{35}   &
f_{36}&=\epsilon ^4 \, \top{36}  \\
f_{37}&= \epsilon ^4 \, \frac{ (1-2 \epsilon ) } {1-\epsilon } \top{37}   &
f_{38}&=\epsilon ^5 \, \top{38}   &
f_{39}&=\epsilon ^5 \, \top{39}  \\
f_{40}&=\epsilon ^5 \, \top{40}   &
f_{41}&=\epsilon ^4\, \top{41}  &
f_{42}&=\epsilon ^5 \, \top{42}  \\
f_{43}&=\epsilon ^4 \, \top{43}   &
f_{44}&=\epsilon ^4 \, \top{44}   &
f_{45}&=\epsilon ^6 \, \top{45}  \\
f_{46}&=\epsilon ^6 \, \top{46}   &
f_{47}&=\epsilon ^5 \, \top{47}   &
f_{48}&=\epsilon ^6 \, \top{48}  \\
f_{49}&=\epsilon ^6 \, \top{49}   &
f_{50}&=\epsilon ^4 \, (1+ \epsilon ) \, \top{50}   &
f_{51}&=\epsilon ^5 \, \top{51}  \\
f_{52}&= \epsilon ^5 \, (1-2 \epsilon ) \, \top{52}    &
f_{53}&=\epsilon ^6 \, \top{53}   &
f_{54}&=\epsilon ^5 \, \top{54}  \\
f_{55}&=\epsilon ^5 \, \top{55}   &
f_{56}&=\epsilon^4 \, (1+\epsilon) \, \top{56}    &
f_{57}&=\epsilon ^5 \, \top{57}  \\
f_{58}&=\epsilon ^6 \, \top{58}   &
f_{59}&=\epsilon ^4 \, (1+ \epsilon) \, \top{59}    &
f_{60}&=\epsilon ^6 \, \top{60}  \\
f_{61}&=\epsilon ^5 \, \top{61}   &
f_{62}&=\epsilon ^5 \, \top{62}   &
f_{63}&=\epsilon ^5 \, \top{63}  \\
f_{64}&=\epsilon ^5 \, \top{64}   &
f_{65}&=\epsilon^5 \, (1-2 \epsilon )  \, \top{65}    &
f_{66}&=\epsilon ^6 \, \top{66}  \\
f_{67}&=\epsilon ^5 \, \top{67}   &
f_{68}&=\epsilon ^6 \, \top{68}   &
f_{69}&=\epsilon ^5 \, \top{69}  \\
f_{70}&= \epsilon ^5 \, (1-2 \epsilon ) \, \top{70}   &
f_{71}&=\epsilon ^5 \, \top{71}  &
f_{72}&=\epsilon ^6 \, \top{72}  \\
f_{73}&=\epsilon ^5 \, \top{73}   &
f_{74}&=\epsilon ^6 \, \top{74}   &
f_{75}&=\epsilon ^5\, \top{75}  \\
f_{76}&=\epsilon ^5 \, \top{76}    &
f_{77}&=\epsilon ^5 \, \top{77}  &
f_{78}&=\epsilon ^5\, (1-2 \epsilon )  \, \top{78}   \\
f_{79}&=\epsilon ^5 \, \top{79}   &
f_{80}&=\epsilon^6 \, \top{80}   &
f_{81}&=\epsilon ^5 \, (1-2 \epsilon )  \, \top{81}   \\
f_{82}&=\epsilon ^6 \, \top{82}   &
f_{83}&=\epsilon ^6 \, \top{83}   &
f_{84}&=\epsilon ^6 \, \top{84}  \\
f_{85}&=\epsilon ^6 \, \top{85}  \stepcounter{equation}\tag{\theequation}\label{def:3loopBasisMIs}
\end{align*}
where the integrals $\top{i}$ are depicted in Figs.~\ref{Fig:3loopMIs1}--\ref{Fig:3loopMIs3}.
As before, the choice of $\{ f \}_{i=1,\ldots,85}$ is motivated by them obeying a $\e$-linear system
  of differential equations in $x$ and $y$. 
}

In this case, the canonical transformation $B$, generically defined in~\eqref{eq:genericBmatrix}, reduces to 
$
B \equiv e^{\magnus{A_{m^2\!,0}}}\, e^{\magnus{D^{[0]}_{x,0}}} \,
e^{\magnus{D^{[1]}_{y,0}}} \,
e^{\magnus{N^{[2]}_{x,0}}}
$,
yielding the canonical basis $\{ g \}_{i=1,\ldots,85}$
{\allowdisplaybreaks[1]
\begin{align*}
g_1&=s \, f_1&
g_2&=t \, f_2&
g_3&=m^2 \, f_3 &
g_4&=s^2 \, f_4\\
g_5&=s \, f_5&
g_6&=s \, f_6&
g_7&=s \, f_7&
g_8&=s \, f_8\\
g_9&= m^2 \, s \, f_9&
g_{10}&=m^2   s \, f_{10}&
g_{11}&= \lambda_s \, f_{11}&
g_{12}&=\lambda_s \, f_{12} \\
g_{13}&=\lambda_t \, f_{13} &
g_{14}&=m^2 \, \left( s \, f_{14}-4 f_{15} \right) &
g_{15}&=\lambda_s \, f_{15}  &
g_{16}&=\lambda_s \, f_{16}  \\
g_{17}&=\lambda_t \, f_{17} &
g_{18}&=s^3 \, f_{18}  &
g_{19}&=s \, f_{19} &
g_{20}&=s^2 \, f_{20} \\
g_{21}&=m^2 \, s^2 \, f_{21}   &
g_{22}&=s \, \lambda_s \, f_{22} &
g_{23}&=s \, \lambda_s \,  f_{23}  &
g_{24}&= \lambda_s \, f_{24}  \\
g_{25}&=m^2 \lambda_s f_{25}  &
g_{26}&=\lambda_s \, f_{26} &
g_{27}&= \lambda_s \, f_{27}  &
g_{28}&=m^2 \, \left(s \, f_{28} -12 f_{27} \right)  \\
g_{29}&=\lambda_s \, f_{29} &
g_{30}&=m^2 \, s \, f_{30} &
g_{31}&=s \, t \, f_{31} &
g_{32}&=u \, f_{32}  \\
g_{33}&=\lambda_u \,f_{33}&
g_{34}&=s \,  t \, f_{34}&
g_{35}&=m^2 \, s \, f_{35} &
g_{36}&=s \, t \,  f_{36} \\
g_{37}&=s \, \left( f_{37} - f_{17}\right)  &
g_{38}&=u \, f_{38}&
g_{39}&=u \, f_{39} &
g_{40}&=\lambda_u \, f_{40} \\
g_{41}&=s \, t \, f_{41}&
g_{42}&=\lambda_u \, f_{42}&
g_{43}&=s \, t \,  f_{43} &
g_{44}&=s \, t \, f_{44} \\
g_{45}&=s \, f_{45} &
g_{46}&=s \, f_{46} &
g_{47}&=s \, \lambda_s \,  f_{47}  &
g_{48}&=\lambda_s \, f_{48}  \\
g_{49}&= \lambda_s \, f_{49}  &
g_{50}&=s \, \lambda_s \,  f_{50}  &
g_{51}&=s  \, \lambda_t \, f_{51} &
g_{52}&=\lambda_s \, f_{52}  \\
g_{53}&=\lambda_t \, f_{53}  &
g_{54}&=m^2 \,  s \, f_{54}  &
g_{55}&=s \, \lambda_s \, f_{55}  &
g_{56}&=s \, \lambda_t \, f_{56}  \\
g_{57}&=s \, t \,  f_{57}  &
g_{58}&=\lambda_u \, f_{58}  &
g_{59}&=m^2 \, s \, f_{59}  &
g_{60}&=t  \, f_{60}  \\
g_{61}&= s \, \lambda_s \, f_{61}  &
g_{62}&=s \, \lambda_t \, f_{62}  &
g_{63}&=s \, \lambda_t \, f_{63}  &
g_{64}&=s^2 \, t \, f_{64}  \\
g_{65}&=s \, \lambda_s \, f_{65}  &
g_{66}&=s \, u \, f_{66}   &
g_{67}&=s^2 \, t \,  f_{67}  &
g_{68}&= s \, u \, f_{68}    \\
g_{69}&=s^2 \,  t \, f_{69} &
g_{70}&=s \, \lambda_s \, f_{70}  &
g_{71}&=s^2 \, t \,  f_{71} &
g_{72}&=s \, u \, f_{72} \\
g_{73}&=s^2 \, t \, f_{73} &
g_{74}&=s \, \lambda_u \, f_{74} &
g_{75}&=m^2 \, s^2 \, f_{75}  &
g_{76}&=m^2 \, s \, t \, f_{76}  \\
g_{77}&=s^2 \, t \, f_{77}  &
g_{78}&=s^2 \, f_{78}  &
g_{79}&=s^2 \, t \, f_{79} &
g_{80}&=s^2 \, \lambda_t  \, f_{80}  \\
g_{81}&=\lefteqn{m^2 \, \Bigl(- 2 f_{24}  -3 f_{26}  +4 f_{27} -f_{29} +2 s \,  f_{47}
                    -2 f_{49}  - 2 \frac{s \, t} {m^2} f_{74} + \frac{s^2}{m^2} f_{81} - 2 s \,  f_{82}  \Bigr)}  \\
g_{82}&=s \, \lambda_s  \,f_{82}  &
g_{83}&=s^3 \, t \, f_{83}  &
g_{84}&=s^2 \, \lambda_s \, f_{84}  &
g_{85}&=s^2 \, \lambda_s \, f_{85}  \stepcounter{equation}\tag{\theequation}\label{def:3loopCanonicalMIs}
\end{align*}
}%
with 
$\lambda_a$ defined below~\eqref{def:2loopCanonicalMIs}.
The sparse matrices $M_i$ $(i=1,\ldots,6)$ appearing in the corresponding canonical
system~\eqref{eq:cansys2} and~\eqref{eq:GenCanSys} are given in Appendix~\ref{App:3loopMatrices}.

%% file: boundaries.tex
\section{Boundary conditions}
\label{sec:boundaries}
The generic solutions~\eqref{eq:sol:gen:Dyson} of the canonical systems at two-
and three-loop are written in terms of $G$-polylogarithms and constants to
be fixed by boundary conditions. The alphabet~\eqref{def:alphabet}
determines the thresholds which appear in the final result. 
Out of six thresholds only two are physical, since they correspond to
the production of massless particles in $s$- and $t$-channels
at $x = 0$ and $y = 0$. 
Imposing the regularity of the generic solutions at the unphysical
thresholds, namely $x=1$, $y=1$, $y=-x$, $y=1-x$, amounts to ruling out the terms that give rise to divergent
behaviours, hence enforcing conditions that unequivocally fix the arbitrary constants.

\subsection{Relations for one-scale integrals}
In general, homogeneous differential equations for single scale
integrals carry only information on the scaling behaviour of the
solution. Boundary constants for such differential equations require the
evaluation of the integrals themselves by independent methods.
Within a multi-scale problem, such as the one we are considering, integrals may depend on more than
one external invariant, and single-scale integrals participate in the regularity conditions of the multi-scale ones.
Therefore, these relations can be exploited to determine the arbitrary constants of the single-scale
integrals. Alternatively, they can reduce the number of independent
single-scale integrals that needs to be independently provided.
Therefore, solving multi-scale systems of differential equations
yields the simultaneous determination of single- and multi-scale MI's,
which are finally expressed in terms of a few single-scale MI's, to be
considered as external {\it input}.

Let us discuss, as a pedagogical example, the systems of DE's for the two-loop master integrals $g_{2}$, $g_{5}$, $g_{8}$
and $g_{11}$, represented by the corresponding topologies in
Fig.~\ref{Fig:2loopMIs}, which read,
\begin{align}
\der{x} \parbox{12mm}{ \includegraphics[width=0.08\textwidth]{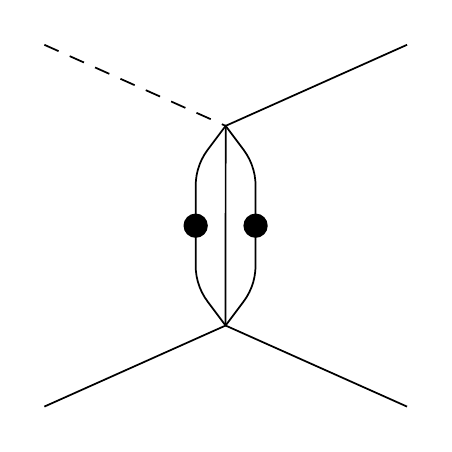}}= {} &  0 \,,\\
\der{x} \parbox{12mm}{ \includegraphics[width=0.08\textwidth]{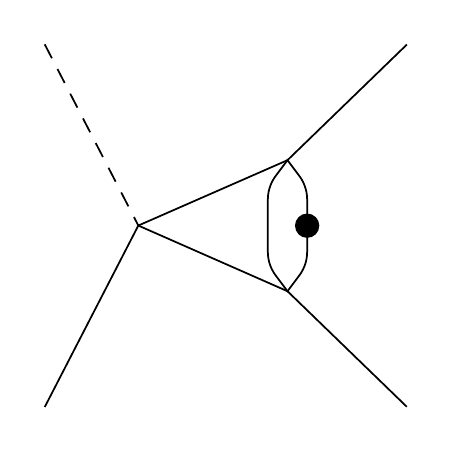}}= {} &  -\frac{2}{x} \parbox{12mm}{ \includegraphics[width=0.08\textwidth]{figures/MI5.pdf}} \,,\\
\der{x} \parbox{12mm}{ \includegraphics[width=0.08\textwidth]{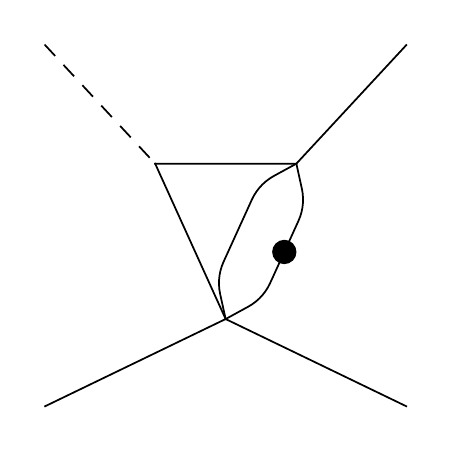}}= {} & 0 \,,\\
\label{boxx}
\begin{split}
\der{x} \parbox{12mm}{ \includegraphics[width=0.08\textwidth]{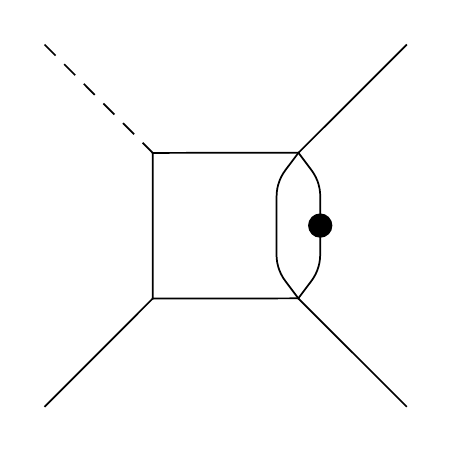}}= {} &
- \frac{3}{2} \frac{1}{1-x} \parbox{12mm}{ \includegraphics[width=0.08\textwidth]{figures/MI2.pdf}}
+ 3 \left( -\frac{1}{1-x} + \frac{1} {1-x-y} \right)  \parbox{12mm}{ \includegraphics[width=0.08\textwidth]{figures/MI5.pdf}}  \\
 +& 3 \left( \frac{1}{1-x} -  \frac{1} {1-x-y} \right)
  \parbox{11mm}{ \includegraphics[width=0.08\textwidth]{figures/MI8.pdf}}
 - \left( \frac{1}{1-x} + \frac{2}{x} +  \frac{1} {1-x-y} \right) 
 \parbox{12mm}{ \includegraphics[width=0.08\textwidth]{figures/MI11.pdf}} \,, 
\end{split}  
\shortintertext{and}
%
\der{y} \parbox{12mm}{ \includegraphics[width=0.08\textwidth]{figures/MI2.pdf}}= {} &  -\frac{2}{y} \parbox{12mm}{ \includegraphics[width=0.08\textwidth]{figures/MI2.pdf}} \,,\\
\der{y} \parbox{12mm}{ \includegraphics[width=0.08\textwidth]{figures/MI5.pdf}}= {} &0 \,,\\
\der{y} \parbox{12mm}{ \includegraphics[width=0.08\textwidth]{figures/MI8.pdf}}= {} & -\frac{1}{2y} \parbox{12mm}{ \includegraphics[width=0.08\textwidth]{figures/MI2.pdf}}+  \frac{1}{1-y} \parbox{12mm}{ \includegraphics[width=0.08\textwidth]{figures/MI8.pdf}} \,,\\
\begin{split}
\der{y} \parbox{12mm}{ \includegraphics[width=0.08\textwidth]{figures/MI11.pdf}} = {} & 
\frac{3} {1-x-y}  \parbox{12mm}{ \includegraphics[width=0.08\textwidth]{figures/MI5.pdf}}
+ 3 \left( \frac{1}{1-y} -  \frac{1} {1-x-y} \right)
  \parbox{12mm}{ \includegraphics[width=0.08\textwidth]{figures/MI8.pdf}} \\
& + \left(-\frac{2}{y} -  \frac{1} {1-x-y} \right)
  \parbox{12mm}{ \includegraphics[width=0.08\textwidth]{figures/MI11.pdf}} \label{boxy} \, .
\end{split}
\end{align}

\noindent
From the regular behavior of (\ref{boxx}) and  (\ref{boxy}) at $(1-x) \to 0$, $(1-y) \to 0$ and $(1-x-y) \to
0$, the following relations can be established:
\begin{eqnarray}
\left( -\frac{3}{2}   \left. \parbox{12mm}{ \includegraphics[width=0.08\textwidth]{figures/MI2.pdf}}  
-3  \parbox{12mm}{ \includegraphics[width=0.08\textwidth]{figures/MI5.pdf}}  
+3 \parbox{12mm}{ \includegraphics[width=0.08\textwidth]{figures/MI8.pdf}} 
- \parbox{12mm}{ \includegraphics[width=0.08\textwidth]{figures/MI11.pdf}}
\right) \right\vert_{x \rightarrow 1} &=&0 \\ 
 \left. \parbox{12mm}{ \includegraphics[width=0.08\textwidth]{figures/MI8.pdf}}
 \right\vert_{y \rightarrow 1}&=&0  \\
 \left( 3 \left. \parbox{12mm}{ \includegraphics[width=0.08\textwidth]{figures/MI5.pdf}}
  -3  \parbox{12mm}{ \includegraphics[width=0.08\textwidth]{figures/MI8.pdf}} 
-  \parbox{12mm}{ \includegraphics[width=0.08\textwidth]{figures/MI11.pdf}}
\right)  \right\vert_{y \rightarrow 1-x} &=&0 
\label{limits}
\end{eqnarray}

The zeroth order term of the $\eps$-expansion is independent of $x$ and $y$.
Therefore, we can combine the equations above in order to find a
relation between (the constant terms of) two one-scale integrals,
\begin{align}
\left. \parbox{12mm}{ \includegraphics[width=0.08\textwidth]{figures/MI5.pdf}} \right\vert_{ \epsilon^0, \, \text{const.}} 
&= \left. - \frac{1}{4} \parbox{12mm}{ \includegraphics[width=0.08\textwidth]{figures/MI2.pdf}}  \right\vert_{ \epsilon^0, \, \text{const.}}
\intertext{%
At higher order in $\eps$, these equations acquire a richer structure,
because constants coming from the limiting values of the
$G$-polylogarithms appear.
For the considered example, the relation at the first order in $\eps$
is unaltered (the only constant which could appear being imaginary, hence not allowed in the Euclidean region),
}
\left. \parbox{12mm}{ \includegraphics[width=0.08\textwidth]{figures/MI5.pdf}}
\right\vert_{ \epsilon^1, \, \text{const.}} &= \left. -
  \frac{1}{4} \parbox{12mm}{ \includegraphics[width=0.08\textwidth]{figures/MI2.pdf}}
\right\vert_{ \epsilon^1, \, \text{const.}} ,
\intertext{but at the second order in $\eps$ the relation becomes,}
\left. \parbox{12mm}{ \includegraphics[width=0.08\textwidth]{figures/MI5.pdf}}
\right\vert_{ \epsilon^2, \, \text{const.}} &= \left. -
  \frac{1}{4} \parbox{12mm}{ \includegraphics[width=0.08\textwidth]{figures/MI2.pdf}}
\right\vert_{ \epsilon^2, \, \text{const.}} -
\left. \frac{\zeta_2}{2}  \parbox{12mm}{ \includegraphics[width=0.08\textwidth]{figures/MI2.pdf}}
\right\vert_{ \epsilon^2, \, \text{const.}} . 
\end{align}

Similar relations are systematically established, so that all MI's can 
be finally determined by providing few simple integrals as external inputs.
At two-loop, the only external input is
$g_{3}$ in \eqref{def:2loopCanonicalMIs}, which can be independently computed and is given by,
\begin{align}
g_{3} &= \eps^2 \frac{\Gamma^2 (1-2 \epsilon ) \, \Gamma^2 (-\epsilon ) \, \Gamma (1+2 \epsilon)}{\Gamma (1-3 \epsilon ) \, \Gamma^3 (1-\epsilon ) \, \Gamma^2 (1+ \epsilon)} ,
\intertext{%
while $g_{3}$ and $g_{9}$ in \eqref{def:3loopCanonicalMIs} are the input integrals for the three-loop MI's, amounting to
}
g_{3} &=\eps^3 \frac{\Gamma^3 (1-2 \epsilon ) \, \Gamma^3 (-\epsilon ) \, \Gamma (1+3 \epsilon )}{\Gamma (1-4 \epsilon ) \, \Gamma^5 (1-\epsilon ) \, \Gamma^3 (1+ \epsilon)} , \\
g_{9} &=\eps^3 \frac{\Gamma^2 (1-2 \epsilon ) \, \Gamma^3 (-\epsilon ) \, \Gamma (1+ 2 \epsilon)}{\Gamma (1-3 \epsilon ) \, \Gamma^4 (1-\epsilon ) \, \Gamma^2 (1+ \epsilon)} \, x^{- \eps} ,
\end{align}
where we omit the common normalization factors~\eqref{eq:intmeasure}.

We would like to observe that the relations between single-scale
integrals, coming from the regularity conditions of multi-scale ones,
seem not to belong to the set of IBP identities needed to derive the
considered systems of differential equations. In the future, it is
worth to investigate whether such relations are truly independent from IBP identities,
or if they would arise when considering larger sets of
identities for increasing powers of denominators and irreducible
scalar products. \\

The sets of 18 two-loop MI's and of 85 three-loop MI's
are respectively collected in two ancillary files, {\tt <2loopMIs.m>} and 
{\tt <3loopMIs.m>}.
In Appendix \ref{App:3loopg83}, we present only the analytic expression of the three
loop $g_{83}$-integral, see Fig.~\ref{Fig:3loopMIs3}, that we consider the representative diagram of this work.

%% file: conclusions.tex
\section{Conclusions}
In this article we presented the analytic expressions of the 85 master
integrals (MI's) of the three-loop ladder-box topology with one massive leg.
Their calculation was performed with the method of differential
equations, namely by solving a system of first order
differential equations fulfilled by the MI's. 
The generic solution of the system was obtained in a purely algebraic way, by means of 
Magnus exponential method, 
and cast in terms of repeated integrations, according to Dyson series
expansion, as recently proposed in Ref.~\cite{Argeri:2014qva}.
The boundary conditions were provided by the regularity of the
solutions at pseudothresholds.

The results of the considered four-leg integrals, as well as of the
tower of three- and two-leg master integrals associated to
subtopologies (including previously unknown two-scale vertex diagrams),
were written as a Taylor expansion in the dimensional regulator parameter
$\epsilon = (4-d)/2$. 
The coefficients of the series are expressed in terms of uniform weight
combinations of multiple polylogarithms and transcendental constants 
up to weight six. 

The considered integrals contribute to the N$^3$LO virtual corrections 
to scattering processes like the three-jet production mediated by vector boson decay, $V^{*} \to jjj$, 
as well as the Higgs plus one-jet production in gluon fusion, $pp \to
Hj$, and to the three-loop one-particle splitting amplitudes.

%% file: 2loopMatrices.tex
\section{Canonical Matrices at Two-Loop}
\label{App:2loopMatrices}
Here we present the sparse matrices $M_i$ $(i=1,\ldots,6)$ appearing in the canonical
system defined in~\eqref{eq:cansys2} and~\eqref{eq:GenCanSys} obeyed by the MI's~\eqref{def:2loopCanonicalMIs}:
\begin{eqnarray}
M_1=\left(
\scalemath{0.9}{
\begin{array}{cccccccccccccccccc}
 -2 & 0 & 0 & 0 & 0 & 0 & 0 & 0 & 0 & 0 & 0 & 0 & 0 & 0 & 0 & 0 & 0 & 0 \\
 0 & 0 & 0 & 0 & 0 & 0 & 0 & 0 & 0 & 0 & 0 & 0 & 0 & 0 & 0 & 0 & 0 & 0 \\
 0 & 0 & 0 & 0 & 0 & 0 & 0 & 0 & 0 & 0 & 0 & 0 & 0 & 0 & 0 & 0 & 0 & 0 \\
 0 & 0 & 0 & -2 & 0 & 0 & 0 & 0 & 0 & 0 & 0 & 0 & 0 & 0 & 0 & 0 & 0 & 0 \\
 0 & 0 & 0 & 0 & -2 & 0 & 0 & 0 & 0 & 0 & 0 & 0 & 0 & 0 & 0 & 0 & 0 & 0 \\
 0 & 0 & 0 & 0 & 0 & -1 & 0 & 0 & 0 & 0 & 0 & 0 & 0 & 0 & 0 & 0 & 0 & 0 \\
 -\frac{1}{2} & 0 & 0 & 0 & 0 & 0 & 0 & 0 & 0 & 0 & 0 & 0 & 0 & 0 & 0 & 0 & 0 & 0 \\
 0 & 0 & 0 & 0 & 0 & 0 & 0 & 0 & 0 & 0 & 0 & 0 & 0 & 0 & 0 & 0 & 0 & 0 \\
 0 & 0 & -\frac{1}{2} & 0 & 0 & 0 & 0 & 0 & -1 & 0 & 0 & 0 & 0 & 0 & 0 & 0 & 0 & 0 \\
 \frac{1}{2} & 0 & \frac{1}{2} & 0 & 0 & -1 & -2 & 0 & 2 & -2 & 0 & 0 & 0 & 0 & 0 & 0 & 0 & 0 \\
 0 & 0 & 0 & 0 & 0 & 0 & 0 & 0 & 0 & 0 & -2 & 0 & 0 & 0 & 0 & 0 & 0 & 0 \\
 0 & -\frac{1}{2} & \frac{1}{2} & 0 & 0 & 0 & 0 & 0 & 0 & 0 & 0 & -2 & 0 & 0 & 0 & 0 & 0 & 0 \\
 \frac{1}{2} & -\frac{1}{2} & 0 & 0 & 0 & 0 & -2 & 2 & 0 & 0 & 0 & 0 & 2 & 0 & 0 & 0 & 0 & 0 \\
 0 & 0 & 0 & 0 & 0 & 0 & 0 & 0 & 0 & 0 & 0 & 0 & 0 & -2 & 0 & 0 & 0 & 0 \\
 0 & 0 & 0 & 0 & 0 & 0 & 0 & 0 & 0 & 0 & 0 & 0 & 0 & 0 & -1 & 0 & 0 & 0 \\
 -\frac{3}{8} & \frac{3}{8} & -\frac{3}{4} & 0 & 0 & 1 & \frac{5}{2} & -\frac{3}{2} & -2 & 1 & 0 & 0 & \frac{3}{2} & \frac{1}{4} & 0 & -1 & 0
   & 0 \\
 0 & 0 & 0 & 0 & 0 & 0 & 0 & 0 & 0 & 0 & 0 & 0 & 0 & 0 & 0 & 0 & -2 & 0 \\
 -\frac{3}{8} & \frac{3}{8} & -\frac{3}{4} & 0 & 0 & 1 & \frac{5}{2} & -\frac{3}{2} & -2 & 1 & 0 & 0 & \frac{3}{2} & \frac{1}{4} & -1 & 1 & 0
   & -2 \\
\end{array}
}
\right)
\end{eqnarray}

\begin{eqnarray}
M_2=\left(
\scalemath{0.9}{
\begin{array}{cccccccccccccccccc}
 0 & 0 & 0 & 0 & 0 & 0 & 0 & 0 & 0 & 0 & 0 & 0 & 0 & 0 & 0 & 0 & 0 & 0 \\
 0 & 0 & 0 & 0 & 0 & 0 & 0 & 0 & 0 & 0 & 0 & 0 & 0 & 0 & 0 & 0 & 0 & 0 \\
 0 & 0 & 0 & 0 & 0 & 0 & 0 & 0 & 0 & 0 & 0 & 0 & 0 & 0 & 0 & 0 & 0 & 0 \\
 0 & 0 & 0 & 0 & 0 & 0 & 0 & 0 & 0 & 0 & 0 & 0 & 0 & 0 & 0 & 0 & 0 & 0 \\
 0 & 0 & 0 & 0 & 0 & 0 & 0 & 0 & 0 & 0 & 0 & 0 & 0 & 0 & 0 & 0 & 0 & 0 \\
 0 & 0 & 0 & 0 & 0 & 0 & 0 & 0 & 0 & 0 & 0 & 0 & 0 & 0 & 0 & 0 & 0 & 0 \\
 0 & 0 & 0 & 0 & 0 & 0 & -1 & 0 & 0 & 0 & 0 & 0 & 0 & 0 & 0 & 0 & 0 & 0 \\
 0 & 0 & 0 & 0 & 0 & 0 & 0 & 0 & 0 & 0 & 0 & 0 & 0 & 0 & 0 & 0 & 0 & 0 \\
 0 & 0 & 0 & 0 & 0 & 0 & 0 & 0 & -1 & 0 & 0 & 0 & 0 & 0 & 0 & 0 & 0 & 0 \\
 0 & 0 & 0 & 0 & 0 & 0 & 0 & 0 & 0 & 2 & 0 & 0 & 0 & 0 & 0 & 0 & 0 & 0 \\
 0 & \frac{3}{2} & 0 & 0 & 0 & 0 & 0 & 0 & 0 & 0 & 2 & 0 & 0 & 0 & 0 & 0 & 0 & 0 \\
 \frac{1}{2} & 0 & -\frac{1}{2} & 0 & 0 & 0 & 0 & 0 & 0 & 0 & 0 & 2 & 0 & 0 & 0 & 0 & 0 & 0 \\
 -\frac{1}{8} & -\frac{1}{8} & 0 & 0 & 0 & 0 & -\frac{1}{2} & -\frac{1}{2} & 0 & 0 & 0 & 0 & -\frac{1}{2} & -\frac{1}{4} & 0 & 0 & 0 & 0 \\
 \frac{3}{4} & \frac{3}{4} & 0 & 0 & 0 & 0 & -3 & 3 & 0 & 0 & 0 & 0 & 3 & \frac{3}{2} & 0 & 0 & 0 & 0 \\
 0 & \frac{3}{2} & 0 & 0 & 0 & 0 & 0 & 0 & 0 & 0 & 0 & 0 & 0 & 0 & 1 & 0 & 0 & 0 \\
 \frac{3}{8} & -\frac{3}{8} & \frac{3}{4} & 0 & 0 & -1 & -\frac{5}{2} & \frac{3}{2} & 2 & -1 & 0 & 0 & -\frac{3}{2} & -\frac{1}{4} & 0 & 1 & 0
   & 0 \\
 \frac{3}{2} & \frac{3}{2} & -\frac{3}{2} & 0 & 0 & 0 & 0 & 0 & 0 & 0 & 2 & 6 & 0 & -1 & 2 & 0 & 2 & 0 \\
 -\frac{9}{8} & -\frac{15}{8} & \frac{3}{2} & -1 & 3 & 1 & -\frac{3}{2} & -\frac{3}{2} & 0 & -3 & -2 & -9 & \frac{3}{2} & \frac{5}{4} & -2 & 0
   & -1 & -1 \\
\end{array}
}
\right)
\end{eqnarray}

\begin{eqnarray}
M_3=\left(
\scalemath{0.9}{
\begin{array}{cccccccccccccccccc}
 0 & 0 & 0 & 0 & 0 & 0 & 0 & 0 & 0 & 0 & 0 & 0 & 0 & 0 & 0 & 0 & 0 & 0 \\
 0 & -2 & 0 & 0 & 0 & 0 & 0 & 0 & 0 & 0 & 0 & 0 & 0 & 0 & 0 & 0 & 0 & 0 \\
 0 & 0 & 0 & 0 & 0 & 0 & 0 & 0 & 0 & 0 & 0 & 0 & 0 & 0 & 0 & 0 & 0 & 0 \\
 0 & 0 & 0 & 0 & 0 & 0 & 0 & 0 & 0 & 0 & 0 & 0 & 0 & 0 & 0 & 0 & 0 & 0 \\
 0 & 0 & 0 & 0 & 0 & 0 & 0 & 0 & 0 & 0 & 0 & 0 & 0 & 0 & 0 & 0 & 0 & 0 \\
 0 & 0 & 0 & 0 & 0 & 0 & 0 & 0 & 0 & 0 & 0 & 0 & 0 & 0 & 0 & 0 & 0 & 0 \\
 0 & 0 & 0 & 0 & 0 & 0 & 0 & 0 & 0 & 0 & 0 & 0 & 0 & 0 & 0 & 0 & 0 & 0 \\
 0 & -\frac{1}{2} & 0 & 0 & 0 & 0 & 0 & 0 & 0 & 0 & 0 & 0 & 0 & 0 & 0 & 0 & 0 & 0 \\
 0 & 0 & 0 & 0 & 0 & 0 & 0 & 0 & 0 & 0 & 0 & 0 & 0 & 0 & 0 & 0 & 0 & 0 \\
 0 & 0 & 0 & 0 & 0 & 0 & 0 & 0 & 0 & 0 & 0 & 0 & 0 & 0 & 0 & 0 & 0 & 0 \\
 0 & 0 & 0 & 0 & 0 & 0 & 0 & 0 & 0 & 0 & -2 & 0 & 0 & 0 & 0 & 0 & 0 & 0 \\
 -\frac{1}{2} & 0 & \frac{1}{2} & 0 & 0 & 0 & 0 & 0 & 0 & 0 & 0 & -2 & 0 & 0 & 0 & 0 & 0 & 0 \\
 -\frac{1}{2} & \frac{1}{2} & 0 & 0 & 0 & 0 & 2 & -2 & 0 & 0 & 0 & 0 & -2 & 0 & 0 & 0 & 0 & 0 \\
 0 & 0 & 0 & 0 & 0 & 0 & 0 & 0 & 0 & 0 & 0 & 0 & 0 & -2 & 0 & 0 & 0 & 0 \\
 0 & 0 & 0 & 0 & 0 & 0 & 0 & 0 & 0 & 0 & 0 & 0 & 0 & 0 & -2 & 0 & 0 & 0 \\
 \frac{3}{8} & -\frac{3}{8} & \frac{3}{4} & 0 & 0 & -1 & -\frac{1}{2} & \frac{3}{2} & 1 & 1 & 0 & 0 & -\frac{3}{2} & \frac{3}{4} & -1 & 1 & 0
   & 0 \\
 -\frac{3}{2} & -\frac{3}{2} & \frac{3}{2} & 0 & 0 & 0 & 0 & 0 & 0 & 0 & -2 & -6 & 0 & 0 & 0 & 0 & -2 & 0 \\
 \frac{9}{8} & \frac{15}{8} & -\frac{3}{2} & 1 & -3 & -1 & -\frac{1}{2} & \frac{3}{2} & 1 & 1 & 2 & 9 & -\frac{3}{2} & -\frac{5}{4} & 2 & 0 &
   1 & 1 \\
\end{array}
}
\right)
\end{eqnarray}

\begin{eqnarray}
M_4=\left(
\scalemath{1}{
\begin{array}{cccccccccccccccccc}
 0 & 0 & 0 & 0 & 0 & 0 & 0 & 0 & 0 & 0 & 0 & 0 & 0 & 0 & 0 & 0 & 0 & 0 \\
 0 & 0 & 0 & 0 & 0 & 0 & 0 & 0 & 0 & 0 & 0 & 0 & 0 & 0 & 0 & 0 & 0 & 0 \\
 0 & 0 & 0 & 0 & 0 & 0 & 0 & 0 & 0 & 0 & 0 & 0 & 0 & 0 & 0 & 0 & 0 & 0 \\
 0 & 0 & 0 & 0 & 0 & 0 & 0 & 0 & 0 & 0 & 0 & 0 & 0 & 0 & 0 & 0 & 0 & 0 \\
 0 & 0 & 0 & 0 & 0 & 0 & 0 & 0 & 0 & 0 & 0 & 0 & 0 & 0 & 0 & 0 & 0 & 0 \\
 0 & 0 & 0 & 0 & 0 & 0 & 0 & 0 & 0 & 0 & 0 & 0 & 0 & 0 & 0 & 0 & 0 & 0 \\
 0 & 0 & 0 & 0 & 0 & 0 & 0 & 0 & 0 & 0 & 0 & 0 & 0 & 0 & 0 & 0 & 0 & 0 \\
 0 & 0 & 0 & 0 & 0 & 0 & 0 & -1 & 0 & 0 & 0 & 0 & 0 & 0 & 0 & 0 & 0 & 0 \\
 0 & 0 & 0 & 0 & 0 & 0 & 0 & 0 & 0 & 0 & 0 & 0 & 0 & 0 & 0 & 0 & 0 & 0 \\
 0 & 0 & 0 & 0 & 0 & 0 & 0 & 0 & 0 & 0 & 0 & 0 & 0 & 0 & 0 & 0 & 0 & 0 \\
 0 & 0 & 0 & 0 & 0 & 0 & 0 & -3 & 0 & 0 & 0 & 0 & 0 & 0 & 0 & 0 & 0 & 0 \\
 0 & \frac{1}{2} & -\frac{1}{2} & 0 & 0 & 0 & 0 & 0 & 0 & 0 & 0 & 0 & 0 & 0 & 0 & 0 & 0 & 0 \\
 0 & 0 & 0 & 0 & 0 & 0 & 0 & 0 & 0 & 0 & 0 & 0 & 0 & 0 & 0 & 0 & 0 & 0 \\
 0 & 0 & 0 & 0 & 0 & 0 & 0 & -6 & 0 & 0 & 0 & 0 & 0 & 0 & 0 & 0 & 0 & 0 \\
 0 & -\frac{3}{2} & \frac{3}{2} & 0 & 0 & 0 & 0 & 0 & 0 & 0 & 0 & 0 & 0 & 0 & 0 & 0 & 0 & 0 \\
 0 & 0 & 0 & 0 & 0 & 0 & 0 & 0 & 0 & 0 & 0 & 0 & 0 & 0 & 0 & -2 & 0 & 0 \\
 0 & 0 & 0 & 0 & 0 & 0 & 0 & 0 & 0 & 0 & 0 & 0 & 0 & 0 & 0 & -2 & 0 & 0 \\
 0 & 0 & 0 & 0 & 0 & 0 & 0 & 0 & 0 & 0 & 0 & 0 & 0 & 0 & 0 & 0 & 0 & 0 \\
\end{array}
}
\right)
\end{eqnarray}


$(M_5)_{13,13} =  4$ is the only non-vanishing entry of $M_5$.

\begin{eqnarray}
M_6=\left(
\scalemath{0.85}{
\begin{array}{cccccccccccccccccc}
 0 & 0 & 0 & 0 & 0 & 0 & 0 & 0 & 0 & 0 & 0 & 0 & 0 & 0 & 0 & 0 & 0 & 0 \\
 0 & 0 & 0 & 0 & 0 & 0 & 0 & 0 & 0 & 0 & 0 & 0 & 0 & 0 & 0 & 0 & 0 & 0 \\
 0 & 0 & 0 & 0 & 0 & 0 & 0 & 0 & 0 & 0 & 0 & 0 & 0 & 0 & 0 & 0 & 0 & 0 \\
 0 & 0 & 0 & 0 & 0 & 0 & 0 & 0 & 0 & 0 & 0 & 0 & 0 & 0 & 0 & 0 & 0 & 0 \\
 0 & 0 & 0 & 0 & 0 & 0 & 0 & 0 & 0 & 0 & 0 & 0 & 0 & 0 & 0 & 0 & 0 & 0 \\
 0 & 0 & 0 & 0 & 0 & 0 & 0 & 0 & 0 & 0 & 0 & 0 & 0 & 0 & 0 & 0 & 0 & 0 \\
 0 & 0 & 0 & 0 & 0 & 0 & 0 & 0 & 0 & 0 & 0 & 0 & 0 & 0 & 0 & 0 & 0 & 0 \\
 0 & 0 & 0 & 0 & 0 & 0 & 0 & 0 & 0 & 0 & 0 & 0 & 0 & 0 & 0 & 0 & 0 & 0 \\
 0 & 0 & 0 & 0 & 0 & 0 & 0 & 0 & 0 & 0 & 0 & 0 & 0 & 0 & 0 & 0 & 0 & 0 \\
 0 & 0 & 0 & 0 & 0 & 0 & 0 & 0 & 0 & 0 & 0 & 0 & 0 & 0 & 0 & 0 & 0 & 0 \\
 0 & 0 & 0 & 0 & -3 & 0 & 0 & 3 & 0 & 0 & 1 & 0 & 0 & 0 & 0 & 0 & 0 & 0 \\
 0 & 0 & 0 & 0 & 0 & 0 & 0 & 0 & 0 & 0 & 0 & 2 & 0 & 0 & 0 & 0 & 0 & 0 \\
 -\frac{1}{8} & -\frac{1}{8} & 0 & 0 & 0 & 0 & -\frac{1}{2} & -\frac{1}{2} & 0 & 0 & 0 & 0 & -\frac{1}{2} & -\frac{1}{4} & 0 & 0 & 0 & 0 \\
 \frac{3}{4} & \frac{3}{4} & 0 & 0 & 0 & 0 & 3 & 3 & 0 & 0 & 0 & 0 & 3 & \frac{3}{2} & 0 & 0 & 0 & 0 \\
 0 & \frac{3}{2} & 0 & 0 & 0 & 0 & 0 & 0 & 3 & 0 & 0 & 0 & 0 & 0 & 1 & 0 & 0 & 0 \\
 \frac{3}{8} & -\frac{3}{8} & \frac{3}{4} & 0 & 0 & -1 & -\frac{5}{2} & \frac{3}{2} & 2 & -1 & 0 & 0 & -\frac{3}{2} & -\frac{1}{4} & 0 & 1 & 0
   & 0 \\
 3 & \frac{3}{2} & -3 & -2 & 0 & 2 & 2 & 0 & -4 & 2 & 2 & 6 & 0 & -1 & 2 & 0 & 2 & 2 \\
 -\frac{9}{8} & -\frac{15}{8} & \frac{3}{2} & 1 & 3 & -1 & \frac{1}{2} & -\frac{3}{2} & -1 & -1 & -2 & -9 & \frac{3}{2} & \frac{5}{4} & -2 & 0
   & -1 & -1 \\
\end{array}
}
\right)
\end{eqnarray}

%% file: 3loopMatrices.tex
\section{Canonical Matrices at Three-Loop}
\label{App:3loopMatrices}
Here we present the sparse matrices $M_i$ $(i=1,\ldots,6)$ appearing in the canonical
system defined in~\eqref{eq:cansys2} and~\eqref{eq:GenCanSys} obeyed by the MI's~\eqref{def:3loopCanonicalMIs}.
We write each matrix in block form,
\begin{align}
  M_i = \left(
  \begin{array}{cc}
    M_i^a & \quad 0_{43 \times 42} \\
    M_i^b & M_i^c
  \end{array}
  \right)\,,
\end{align}
where $0_{43 \times 42}$ is the null $43\times 42$ matrix.

\input{matrices/3loopM1a}
\input{matrices/3loopM1b}
\input{matrices/3loopM1c}

\input{matrices/3loopM2a}
\input{matrices/3loopM2b}
\input{matrices/3loopM2c}

\input{matrices/3loopM3a}
\input{matrices/3loopM3b}
\input{matrices/3loopM3c}

\input{matrices/3loopM4a}
\input{matrices/3loopM4b}
\input{matrices/3loopM4c}

The matrix $M_5$ is mainly composed of zeroes and we only list its non-vanishing entries:
\input{matrices/3loopM5}

\input{matrices/3loopM6a}
\input{matrices/3loopM6b}
\input{matrices/3loopM6c}

%% file: matrices/3loopM1a.tex
\begin{align}
M_1^{(a)} = \scalemath {0.45}{\left(
\begin{array}{ccccccccccccccccccccccccccccccccccccccccccc}
 -3 & 0 & 0 & 0 & 0 & 0 & 0 & 0 & 0 & 0 & 0 & 0 & 0 & 0 & 0 & 0 & 0 & 0 & 0 & 0 & 0 & 0 & 0 & 0 & 0 & 0 & 0 & 0 & 0 & 0 & 0 & 0 & 0 & 0 & 0 & 0 & 0 & 0 & 0 & 0 & 0 & 0 & 0 \\
 0 & 0 & 0 & 0 & 0 & 0 & 0 & 0 & 0 & 0 & 0 & 0 & 0 & 0 & 0 & 0 & 0 & 0 & 0 & 0 & 0 & 0 & 0 & 0 & 0 & 0 & 0 & 0 & 0 & 0 & 0 & 0 & 0 & 0 & 0 & 0 & 0 & 0 & 0 & 0 & 0 & 0 & 0 \\
 0 & 0 & 0 & 0 & 0 & 0 & 0 & 0 & 0 & 0 & 0 & 0 & 0 & 0 & 0 & 0 & 0 & 0 & 0 & 0 & 0 & 0 & 0 & 0 & 0 & 0 & 0 & 0 & 0 & 0 & 0 & 0 & 0 & 0 & 0 & 0 & 0 & 0 & 0 & 0 & 0 & 0 & 0 \\
 0 & 0 & 0 & -3 & 0 & 0 & 0 & 0 & 0 & 0 & 0 & 0 & 0 & 0 & 0 & 0 & 0 & 0 & 0 & 0 & 0 & 0 & 0 & 0 & 0 & 0 & 0 & 0 & 0 & 0 & 0 & 0 & 0 & 0 & 0 & 0 & 0 & 0 & 0 & 0 & 0 & 0 & 0 \\
 0 & 0 & 0 & 0 & -3 & 0 & 0 & 0 & 0 & 0 & 0 & 0 & 0 & 0 & 0 & 0 & 0 & 0 & 0 & 0 & 0 & 0 & 0 & 0 & 0 & 0 & 0 & 0 & 0 & 0 & 0 & 0 & 0 & 0 & 0 & 0 & 0 & 0 & 0 & 0 & 0 & 0 & 0 \\
 0 & 0 & 0 & 0 & 0 & -3 & 0 & 0 & 0 & 0 & 0 & 0 & 0 & 0 & 0 & 0 & 0 & 0 & 0 & 0 & 0 & 0 & 0 & 0 & 0 & 0 & 0 & 0 & 0 & 0 & 0 & 0 & 0 & 0 & 0 & 0 & 0 & 0 & 0 & 0 & 0 & 0 & 0 \\
 0 & 0 & 0 & 0 & 0 & 0 & -3 & 0 & 0 & 0 & 0 & 0 & 0 & 0 & 0 & 0 & 0 & 0 & 0 & 0 & 0 & 0 & 0 & 0 & 0 & 0 & 0 & 0 & 0 & 0 & 0 & 0 & 0 & 0 & 0 & 0 & 0 & 0 & 0 & 0 & 0 & 0 & 0 \\
 0 & 0 & 0 & 0 & 0 & 0 & 0 & -3 & 0 & 0 & 0 & 0 & 0 & 0 & 0 & 0 & 0 & 0 & 0 & 0 & 0 & 0 & 0 & 0 & 0 & 0 & 0 & 0 & 0 & 0 & 0 & 0 & 0 & 0 & 0 & 0 & 0 & 0 & 0 & 0 & 0 & 0 & 0 \\
\end{array}
\right)}
 \end{align}

%% file: matrices/3loopM1b.tex
\begin{align}
M_1^{(b)} = \scalemath {0.45}{\left(
\begin{array}{ccccccccccccccccccccccccccccccccccccccccccc}
 0 & 0 & 0 & 0 & 0 & 0 & 0 & 0 & 0 & 0 & 0 & 0 & 0 & 0 & 0 & 0 & 0 & 0 & 0 & 0 & 0 & 0 & 0 & 0 & 0 & 0 & 0 & 0 & 0 & 0 & 0 & 0 & 0 & 0 & 0 & 0 & 0 & 0 & 0 & 0 & 0 & 0 & 0 \\
 0 & 0 & 0 & 0 & 0 & 0 & 0 & 0 & 0 & 0 & 0 & 0 & 0 & 0 & 0 & 0 & 0 & 0 & 0 & 0 & 0 & 0 & 0 & 0 & 0 & 0 & 0 & 0 & 0 & 0 & 0 & 0 & 0 & 0 & 0 & 0 & 0 & 0 & 0 & 0 & 0 & 0 & 0 \\
 0 & 0 & 0 & 0 & 0 & 0 & 0 & 0 & 0 & 0 & 0 & 0 & 0 & 0 & 0 & 0 & 0 & 0 & 0 & 0 & 0 & 0 & 0 & 0 & 0 & 0 & 0 & 0 & 0 & 0 & 0 & 0 & 0 & 0 & 0 & 0 & 0 & 0 & 0 & 0 & 0 & 0 & 0 \\
 0 & 0 & 0 & \frac{1}{2} & 0 & 0 & 0 & 0 & \frac{1}{2} & 0 & 0 & 0 & 0 & 0 & 0 & 0 & 0 & 0 & 0 & 0 & -1 & -2 & 2 & 0 & 0 & 0 & 0 & 0 & 0 & 0 & 0 & 0 & 0 & 0 & 0 & 0 & 0 & 0 & 0 & 0 & 0 & 0 & 0 \\
 0 & 0 & 0 & 0 & 0 & 0 & 0 & 0 & 0 & 0 & 0 & 0 & 0 & 0 & 0 & 0 & 0 & 0 & 0 & 0 & 0 & 0 & 0 & 0 & 0 & 0 & 0 & 0 & -1 & 0 & 0 & 0 & 0 & 0 & 0 & 0 & 0 & 0 & 0 & 0 & 0 & 0 & 0 \\
 \frac{5}{36} & 0 & -\frac{1}{12} & -\frac{1}{4} & 1 & 0 & 0 & 0 & 0 & -\frac{1}{4} & 1 & -\frac{9}{8} & 0 & 0 & 0 & -\frac{1}{8} & 0 & 0 & 0 & 0 & 0 & \frac{1}{2} & 0 & \frac{3}{2} & -\frac{1}{2} & 0 & 2 & \frac{1}{3} & 0 & 0 & 0 & 0 & 0 & 0 & 0 & 0 & 0 & 0 & 0 & 0 & 0 & 0 & 0 \\
 1 & 0 & -\frac{1}{3} & -\frac{3}{2} & 4 & 0 & 0 & 0 & 0 & -\frac{3}{2} & -\frac{10}{3} & -\frac{27}{4} & 0 & 0 & 0 & \frac{3}{4} & 0 & 0 & 0 & 0 & 0 & 5 & 0 & -7 & 3 & 0 & 8 & 2 & 0 & 0 & 0 & 0 & 0 & 0 & 0 & 0 & 0 & 0 & 0 & 0 & 0 & 0 & 0 \\
 0 & 0 & 0 & 0 & 0 & 0 & 0 & -\frac{7}{2} & 0 & 0 & 0 & 0 & \frac{3}{4} & 0 & 0 & 0 & -\frac{1}{2} & 0 & 0 & 0 & 0 & 0 & 0 & 0 & 0 & 0 & 0 & 0 & 0 & -\frac{3}{2} & 0 & 0 & -\frac{3}{2} & \frac{1}{2} & \frac{9}{4} & -\frac{1}{2} & -\frac{5}{2} & 0 & 0 & 0 & 0 & 0 & 0 \\
\end{array}
\right)}
 \end{align}

%% file: matrices/3loopM1c.tex
\begin{align}
M_1^{(c)} = \scalemath {0.45}{\left(
\begin{array}{cccccccccccccccccccccccccccccccccccccccccc}
 -1 & 0 & 0 & 0 & 0 & 0 & 0 & 0 & 0 & 0 & 0 & 0 & 0 & 0 & 0 & 0 & 0 & 0 & 0 & 0 & 0 & 0 & 0 & 0 & 0 & 0 & 0 & 0 & 0 & 0 & 0 & 0 & 0 & 0 & 0 & 0 & 0 & 0 & 0 & 0 & 0 & 0 \\
 0 & -3 & 0 & 0 & 0 & 0 & 0 & 0 & 0 & 0 & 0 & 0 & 0 & 0 & 0 & 0 & 0 & 0 & 0 & 0 & 0 & 0 & 0 & 0 & 0 & 0 & 0 & 0 & 0 & 0 & 0 & 0 & 0 & 0 & 0 & 0 & 0 & 0 & 0 & 0 & 0 & 0 \\
 0 & 0 & -3 & 0 & 0 & 0 & 0 & 0 & 0 & 0 & 0 & 0 & 0 & 0 & 0 & 0 & 0 & 0 & 0 & 0 & 0 & 0 & 0 & 0 & 0 & 0 & 0 & 0 & 0 & 0 & 0 & 0 & 0 & 0 & 0 & 0 & 0 & 0 & 0 & 0 & 0 & 0 \\
 0 & 0 & 0 & -3 & 0 & 0 & 0 & 0 & 0 & 0 & 0 & 0 & 0 & 0 & 0 & 0 & 0 & 0 & 0 & 0 & 0 & 0 & 0 & 0 & 0 & 0 & 0 & 0 & 0 & 0 & 0 & 0 & 0 & 0 & 0 & 0 & 0 & 0 & 0 & 0 & 0 & 0 \\
 0 & 0 & 0 & 0 & -3 & 0 & 0 & 0 & 0 & 0 & 0 & 0 & 0 & 0 & 0 & 0 & 0 & 0 & 0 & 0 & 0 & 0 & 0 & 0 & 0 & 0 & 0 & 0 & 0 & 0 & 0 & 0 & 0 & 0 & 0 & 0 & 0 & 0 & 0 & 0 & 0 & 0 \\
 0 & 0 & 0 & 0 & 0 & -\frac{1}{2} & -\frac{3}{4} & 0 & 0 & 0 & 0 & 0 & 0 & 0 & 0 & 0 & 0 & 0 & 0 & 0 & 0 & 0 & 0 & 0 & 0 & 0 & 0 & 0 & 0 & 0 & 0 & 0 & 0 & 0 & 0 & 0 & 0 & 0 & 0 & 0 & 0 & 0 \\
 0 & 0 & 0 & 0 & 0 & 5 & -\frac{9}{2} & 0 & 0 & 0 & 0 & 0 & 0 & 0 & 0 & 0 & 0 & 0 & 0 & 0 & 0 & 0 & 0 & 0 & 0 & 0 & 0 & 0 & 0 & 0 & 0 & 0 & 0 & 0 & 0 & 0 & 0 & 0 & 0 & 0 & 0 & 0 \\
 0 & 0 & 0 & 0 & 0 & 0 & 0 & -\frac{3}{2} & -\frac{3}{2} & 0 & 0 & 0 & 0 & 0 & 0 & 0 & 0 & 0 & 0 & 0 & 0 & 0 & 0 & 0 & 0 & 0 & 0 & 0 & 0 & 0 & 0 & 0 & 0 & 0 & 0 & 0 & 0 & 0 & 0 & 0 & 0 & 0 \\
\end{array}
\right)}
 \end{align}

%% file: matrices/3loopM2a.tex
\begin{align}
M_2^{(a)} = \scalemath {0.45}{\left(
\begin{array}{ccccccccccccccccccccccccccccccccccccccccccc}
 0 & 0 & 0 & 0 & 0 & 0 & 0 & 0 & 0 & 0 & 0 & 0 & 0 & 0 & 0 & 0 & 0 & 0 & 0 & 0 & 0 & 0 & 0 & 0 & 0 & 0 & 0 & 0 & 0 & 0 & 0 & 0 & 0 & 0 & 0 & 0 & 0 & 0 & 0 & 0 & 0 & 0 & 0 \\
 0 & 0 & 0 & 0 & 0 & 0 & 0 & 0 & 0 & 0 & 0 & 0 & 0 & 0 & 0 & 0 & 0 & 0 & 0 & 0 & 0 & 0 & 0 & 0 & 0 & 0 & 0 & 0 & 0 & 0 & 0 & 0 & 0 & 0 & 0 & 0 & 0 & 0 & 0 & 0 & 0 & 0 & 0 \\
 0 & 0 & 0 & 0 & 0 & 0 & 0 & 0 & 0 & 0 & 0 & 0 & 0 & 0 & 0 & 0 & 0 & 0 & 0 & 0 & 0 & 0 & 0 & 0 & 0 & 0 & 0 & 0 & 0 & 0 & 0 & 0 & 0 & 0 & 0 & 0 & 0 & 0 & 0 & 0 & 0 & 0 & 0 \\
 0 & 0 & 0 & 0 & 0 & 0 & 0 & 0 & 0 & 0 & 0 & 0 & 0 & 0 & 0 & 0 & 0 & 0 & 0 & 0 & 0 & 0 & 0 & 0 & 0 & 0 & 0 & 0 & 0 & 0 & 0 & 0 & 0 & 0 & 0 & 0 & 0 & 0 & 0 & 0 & 0 & 0 & 0 \\
 0 & 0 & 0 & 0 & 0 & 0 & 0 & 0 & 0 & 0 & 0 & 0 & 0 & 0 & 0 & 0 & 0 & 0 & 0 & 0 & 0 & 0 & 0 & 0 & 0 & 0 & 0 & 0 & 0 & 0 & 0 & 0 & 0 & 0 & 0 & 0 & 0 & 0 & 0 & 0 & 0 & 0 & 0 \\
 0 & 0 & 0 & 0 & 0 & 0 & 0 & 0 & 0 & 0 & 0 & 0 & 0 & 0 & 0 & 0 & 0 & 0 & 0 & 0 & 0 & 0 & 0 & 0 & 0 & 0 & 0 & 0 & 0 & 0 & 0 & 0 & 0 & 0 & 0 & 0 & 0 & 0 & 0 & 0 & 0 & 0 & 0 \\
 0 & 0 & 0 & 0 & 0 & 0 & 0 & 0 & 0 & 0 & 0 & 0 & 0 & 0 & 0 & 0 & 0 & 0 & 0 & 0 & 0 & 0 & 0 & 0 & 0 & 0 & 0 & 0 & 0 & 0 & 0 & 0 & 0 & 0 & 0 & 0 & 0 & 0 & 0 & 0 & 0 & 0 & 0 \\
 0 & 0 & 0 & 0 & 0 & 0 & 0 & 0 & 0 & 0 & 0 & 0 & 0 & 0 & 0 & 0 & 0 & 0 & 0 & 0 & 0 & 0 & 0 & 0 & 0 & 0 & 0 & 0 & 0 & 0 & 0 & 0 & 0 & 0 & 0 & 0 & 0 & 0 & 0 & 0 & 0 & 0 & 0 \\
\end{array}
\right)}
 \end{align}

%% file: matrices/3loopM2b.tex
\begin{align}
M_2^{(b)} = \scalemath {0.45}{\left(
\begin{array}{ccccccccccccccccccccccccccccccccccccccccccc}
 0 & \frac{4}{3} & 0 & 0 & 0 & 0 & 0 & 0 & 0 & 0 & 0 & 0 & 0 & 0 & 0 & 0 & 0 & 0 & 0 & 0 & 0 & 0 & 0 & 0 & 0 & 0 & 0 & 0 & 0 & 0 & 0 & 0 & 0 & 0 & 0 & 0 & 0 & 0 & 0 & 0 & 0 & 0 & 0 \\
 0 & 0 & 0 & 0 & 0 & 0 & 0 & 0 & 0 & 0 & 0 & 0 & 0 & 0 & 0 & 0 & 0 & 0 & 0 & 0 & 0 & 0 & 0 & 0 & 0 & 0 & 0 & 0 & 0 & 0 & 0 & 0 & 0 & 0 & 0 & 0 & 0 & 0 & 0 & 0 & 0 & 0 & 0 \\
 0 & 0 & 0 & 0 & 0 & 0 & 0 & 0 & 0 & 0 & 0 & 0 & 0 & 0 & 0 & 0 & 0 & 0 & 0 & 0 & 0 & 0 & 0 & 0 & 0 & 0 & 0 & 0 & 0 & 0 & 0 & 0 & 0 & 0 & 0 & 0 & 0 & 0 & 0 & 0 & 0 & 0 & 0 \\
 0 & 0 & 0 & 0 & 0 & 0 & 0 & 0 & 0 & 0 & 0 & 0 & 0 & 0 & 0 & 0 & 0 & 0 & 0 & 0 & 0 & 0 & 0 & 0 & 0 & 0 & 0 & 0 & 0 & 0 & 0 & 0 & 0 & 0 & 0 & 0 & 0 & 0 & 0 & 0 & 0 & 0 & 0 \\
 0 & 0 & 0 & 0 & 0 & 0 & 0 & 0 & 0 & 0 & 0 & 0 & 0 & 0 & 0 & 0 & 0 & 0 & 0 & 0 & 0 & 0 & 0 & 0 & 0 & 0 & 0 & 0 & 1 & 0 & 0 & 0 & 0 & 0 & 0 & 0 & 0 & 0 & 0 & 0 & 0 & 0 & 0 \\
 0 & 0 & 0 & 0 & 0 & 0 & 0 & 0 & 0 & 0 & 0 & 0 & 0 & 0 & 0 & 0 & 0 & 0 & 0 & 0 & 0 & 0 & 0 & 0 & 0 & 0 & 0 & 0 & 0 & 0 & 0 & 0 & 0 & 0 & 0 & 0 & 0 & 0 & 0 & 0 & 0 & 0 & 0 \\
 0 & 0 & 0 & 0 & 0 & 0 & 0 & 0 & 0 & 0 & \frac{28}{3} & 0 & 0 & 0 & 0 & 0 & 0 & 0 & 0 & 0 & 0 & -2 & 0 & 16 & -6 & 0 & 4 & 0 & 0 & 0 & 0 & 0 & 0 & 0 & 0 & 0 & 0 & 0 & 0 & 0 & 0 & 0 & 0 \\
 0 & 0 & 0 & 0 & 0 & 0 & 0 & \frac{7}{2} & 0 & 0 & 0 & 0 & -\frac{3}{4} & 0 & 0 & 0 & \frac{1}{2} & 0 & 0 & 0 & 0 & 0 & 0 & 0 & 0 & 0 & 0 & 0 & 0 & \frac{3}{2} & 0 & 0 & \frac{3}{2} & -\frac{1}{2} & -\frac{9}{4} & \frac{1}{2} & \frac{5}{2} & 0 & 0 & 0 & 0 & 0 & 0 \\
\end{array}
\right)}
 \end{align}

%% file: matrices/3loopM2c.tex
\begin{align}
M_2^{(c)} = \scalemath {0.45}{\left(
\begin{array}{cccccccccccccccccccccccccccccccccccccccccc}
 1 & 0 & 0 & 0 & 0 & 0 & 0 & 0 & 0 & 0 & 0 & 0 & 0 & 0 & 0 & 0 & 0 & 0 & 0 & 0 & 0 & 0 & 0 & 0 & 0 & 0 & 0 & 0 & 0 & 0 & 0 & 0 & 0 & 0 & 0 & 0 & 0 & 0 & 0 & 0 & 0 & 0 \\
 0 & 0 & 0 & 0 & 0 & 0 & 0 & 0 & 0 & 0 & 0 & 0 & 0 & 0 & 0 & 0 & 0 & 0 & 0 & 0 & 0 & 0 & 0 & 0 & 0 & 0 & 0 & 0 & 0 & 0 & 0 & 0 & 0 & 0 & 0 & 0 & 0 & 0 & 0 & 0 & 0 & 0 \\
 0 & 0 & 0 & 0 & 0 & 0 & 0 & 0 & 0 & 0 & 0 & 0 & 0 & 0 & 0 & 0 & 0 & 0 & 0 & 0 & 0 & 0 & 0 & 0 & 0 & 0 & 0 & 0 & 0 & 0 & 0 & 0 & 0 & 0 & 0 & 0 & 0 & 0 & 0 & 0 & 0 & 0 \\
 0 & 0 & 0 & 2 & 0 & 0 & 0 & 0 & 0 & 0 & 0 & 0 & 0 & 0 & 0 & 0 & 0 & 0 & 0 & 0 & 0 & 0 & 0 & 0 & 0 & 0 & 0 & 0 & 0 & 0 & 0 & 0 & 0 & 0 & 0 & 0 & 0 & 0 & 0 & 0 & 0 & 0 \\
 0 & 0 & 0 & 0 & 0 & 0 & 0 & 0 & 0 & 0 & 0 & 0 & 0 & 0 & 0 & 0 & 0 & 0 & 0 & 0 & 0 & 0 & 0 & 0 & 0 & 0 & 0 & 0 & 0 & 0 & 0 & 0 & 0 & 0 & 0 & 0 & 0 & 0 & 0 & 0 & 0 & 0 \\
 0 & 0 & 0 & 0 & 0 & 2 & 0 & 0 & 0 & 0 & 0 & 0 & 0 & 0 & 0 & 0 & 0 & 0 & 0 & 0 & 0 & 0 & 0 & 0 & 0 & 0 & 0 & 0 & 0 & 0 & 0 & 0 & 0 & 0 & 0 & 0 & 0 & 0 & 0 & 0 & 0 & 0 \\
 0 & 0 & 0 & 0 & 0 & 16 & -2 & 0 & 0 & 0 & 0 & 0 & 0 & 0 & 0 & 0 & 0 & 0 & 0 & 0 & 0 & 0 & 0 & 0 & 0 & 0 & 0 & 0 & 0 & 0 & 0 & 0 & 0 & 0 & 0 & 0 & 0 & 0 & 0 & 0 & 0 & 0 \\
 0 & 0 & 0 & 0 & 0 & 0 & 0 & \frac{3}{2} & \frac{3}{2} & 0 & 0 & 0 & 0 & 0 & 0 & 0 & 0 & 0 & 0 & 0 & 0 & 0 & 0 & 0 & 0 & 0 & 0 & 0 & 0 & 0 & 0 & 0 & 0 & 0 & 0 & 0 & 0 & 0 & 0 & 0 & 0 & 0 \\
\end{array}
\right)}
 \end{align}

%% file: matrices/3loopM3a.tex
\begin{align}
M_3^{(a)} = \scalemath {0.45}{\left(
\begin{array}{ccccccccccccccccccccccccccccccccccccccccccc}
 0 & 0 & 0 & 0 & 0 & 0 & 0 & 0 & 0 & 0 & 0 & 0 & 0 & 0 & 0 & 0 & 0 & 0 & 0 & 0 & 0 & 0 & 0 & 0 & 0 & 0 & 0 & 0 & 0 & 0 & 0 & 0 & 0 & 0 & 0 & 0 & 0 & 0 & 0 & 0 & 0 & 0 & 0 \\
 0 & -3 & 0 & 0 & 0 & 0 & 0 & 0 & 0 & 0 & 0 & 0 & 0 & 0 & 0 & 0 & 0 & 0 & 0 & 0 & 0 & 0 & 0 & 0 & 0 & 0 & 0 & 0 & 0 & 0 & 0 & 0 & 0 & 0 & 0 & 0 & 0 & 0 & 0 & 0 & 0 & 0 & 0 \\
 0 & 0 & 0 & 0 & 0 & 0 & 0 & 0 & 0 & 0 & 0 & 0 & 0 & 0 & 0 & 0 & 0 & 0 & 0 & 0 & 0 & 0 & 0 & 0 & 0 & 0 & 0 & 0 & 0 & 0 & 0 & 0 & 0 & 0 & 0 & 0 & 0 & 0 & 0 & 0 & 0 & 0 & 0 \\
 0 & 0 & 0 & 0 & 0 & 0 & 0 & 0 & 0 & 0 & 0 & 0 & 0 & 0 & 0 & 0 & 0 & 0 & 0 & 0 & 0 & 0 & 0 & 0 & 0 & 0 & 0 & 0 & 0 & 0 & 0 & 0 & 0 & 0 & 0 & 0 & 0 & 0 & 0 & 0 & 0 & 0 & 0 \\
 0 & 0 & 0 & 0 & 0 & 0 & 0 & 0 & 0 & 0 & 0 & 0 & 0 & 0 & 0 & 0 & 0 & 0 & 0 & 0 & 0 & 0 & 0 & 0 & 0 & 0 & 0 & 0 & 0 & 0 & 0 & 0 & 0 & 0 & 0 & 0 & 0 & 0 & 0 & 0 & 0 & 0 & 0 \\
 0 & 0 & 0 & 0 & 0 & 0 & 0 & 0 & 0 & 0 & 0 & 0 & 0 & 0 & 0 & 0 & 0 & 0 & 0 & 0 & 0 & 0 & 0 & 0 & 0 & 0 & 0 & 0 & 0 & 0 & 0 & 0 & 0 & 0 & 0 & 0 & 0 & 0 & 0 & 0 & 0 & 0 & 0 \\
 0 & 0 & 0 & 0 & 0 & 0 & 0 & 0 & 0 & 0 & 0 & 0 & 0 & 0 & 0 & 0 & 0 & 0 & 0 & 0 & 0 & 0 & 0 & 0 & 0 & 0 & 0 & 0 & 0 & 0 & 0 & 0 & 0 & 0 & 0 & 0 & 0 & 0 & 0 & 0 & 0 & 0 & 0 \\
 0 & 0 & 0 & 0 & 0 & 0 & 0 & 0 & 0 & 0 & 0 & 0 & 0 & 0 & 0 & 0 & 0 & 0 & 0 & 0 & 0 & 0 & 0 & 0 & 0 & 0 & 0 & 0 & 0 & 0 & 0 & 0 & 0 & 0 & 0 & 0 & 0 & 0 & 0 & 0 & 0 & 0 & 0 \\
\end{array}
\right)}
 \end{align}

%% file: matrices/3loopM3b.tex
\begin{align}
M_3^{(b)} = \scalemath {0.45}{\left(
\begin{array}{ccccccccccccccccccccccccccccccccccccccccccc}
 0 & 0 & 0 & 0 & 0 & 0 & 0 & 0 & 0 & 0 & 0 & 0 & 0 & 0 & 0 & 0 & 0 & 0 & 0 & 0 & 0 & 0 & 0 & 0 & 0 & 0 & 0 & 0 & 0 & 0 & 0 & 0 & 0 & 0 & 0 & 0 & 0 & 0 & 0 & 0 & 0 & 0 & 0 \\
 0 & 0 & 0 & 0 & 0 & 0 & 0 & 0 & 0 & 0 & 0 & 0 & 0 & 0 & 0 & 0 & 0 & 0 & 0 & 0 & 0 & 0 & 0 & 0 & 0 & 0 & 0 & 0 & 0 & 0 & 0 & 0 & 0 & 0 & 0 & 0 & 0 & 0 & 0 & 0 & 0 & 0 & 0 \\
 0 & 0 & 0 & 0 & 0 & 0 & 0 & 0 & 0 & 0 & 0 & 0 & 0 & 0 & 0 & 0 & 0 & 0 & 0 & 0 & 0 & 0 & 0 & 0 & 0 & 0 & 0 & 0 & 0 & 0 & 0 & 0 & 0 & 0 & 0 & 0 & 0 & 0 & 0 & 0 & 0 & 0 & 0 \\
 0 & 0 & 0 & 0 & 0 & 0 & 0 & 0 & 0 & 0 & 0 & 0 & 0 & 0 & 0 & 0 & 0 & 0 & 0 & 0 & 0 & 0 & 0 & 0 & 0 & 0 & 0 & 0 & 0 & 0 & 0 & 0 & 0 & 0 & 0 & 0 & 0 & 0 & 0 & 0 & 0 & 0 & 0 \\
 0 & 0 & 0 & 0 & 0 & 0 & 0 & 0 & 0 & 0 & 0 & 0 & 0 & 0 & 0 & 0 & 0 & 0 & 0 & 0 & 0 & 0 & 0 & 0 & 0 & 0 & 0 & 0 & 0 & 0 & 0 & 0 & 0 & 0 & 0 & 0 & 0 & 0 & 0 & 0 & 0 & 0 & 0 \\
 0 & 0 & 0 & 0 & 0 & 0 & 0 & 0 & 0 & 0 & 0 & 0 & 0 & 0 & 0 & 0 & 0 & 0 & 0 & 0 & 0 & 0 & 0 & 0 & 0 & 0 & 0 & 0 & 0 & 0 & 0 & 0 & 0 & 0 & 0 & 0 & 0 & 0 & 0 & 0 & 0 & 0 & 0 \\
 0 & 0 & 0 & 0 & 0 & 0 & 0 & 0 & 0 & 0 & 0 & 0 & 0 & 0 & 0 & 0 & 0 & 0 & 0 & 0 & 0 & 0 & 0 & 0 & 0 & 0 & 0 & 0 & 0 & 0 & 0 & 0 & 0 & 0 & 0 & 0 & 0 & 0 & 0 & 0 & 0 & 0 & 0 \\
 0 & 0 & 0 & 0 & 0 & 0 & 0 & \frac{1}{2} & 0 & 0 & 0 & 0 & \frac{3}{4} & 0 & 0 & 0 & -\frac{1}{2} & 0 & 0 & 0 & 0 & 0 & 0 & 0 & 0 & 0 & 0 & 0 & 0 & \frac{3}{2} & 0 & 0 & -\frac{3}{2} & \frac{1}{2} & -\frac{3}{4} & -\frac{1}{2} & \frac{3}{2} & 0 & 0 & 0 & 0 & 0 & 0 \\
\end{array}
\right)}
 \end{align}

%% file: matrices/3loopM3c.tex
\begin{align}
M_3^{(c)} = \scalemath {0.45}{\left(
\begin{array}{cccccccccccccccccccccccccccccccccccccccccc}
 -3 & 0 & 0 & 0 & 0 & 0 & 0 & 0 & 0 & 0 & 0 & 0 & 0 & 0 & 0 & 0 & 0 & 0 & 0 & 0 & 0 & 0 & 0 & 0 & 0 & 0 & 0 & 0 & 0 & 0 & 0 & 0 & 0 & 0 & 0 & 0 & 0 & 0 & 0 & 0 & 0 & 0 \\
 0 & 0 & 0 & 0 & 0 & 0 & 0 & 0 & 0 & 0 & 0 & 0 & 0 & 0 & 0 & 0 & 0 & 0 & 0 & 0 & 0 & 0 & 0 & 0 & 0 & 0 & 0 & 0 & 0 & 0 & 0 & 0 & 0 & 0 & 0 & 0 & 0 & 0 & 0 & 0 & 0 & 0 \\
 0 & 0 & 0 & 0 & 0 & 0 & 0 & 0 & 0 & 0 & 0 & 0 & 0 & 0 & 0 & 0 & 0 & 0 & 0 & 0 & 0 & 0 & 0 & 0 & 0 & 0 & 0 & 0 & 0 & 0 & 0 & 0 & 0 & 0 & 0 & 0 & 0 & 0 & 0 & 0 & 0 & 0 \\
 0 & 0 & 0 & 0 & 0 & 0 & 0 & 0 & 0 & 0 & 0 & 0 & 0 & 0 & 0 & 0 & 0 & 0 & 0 & 0 & 0 & 0 & 0 & 0 & 0 & 0 & 0 & 0 & 0 & 0 & 0 & 0 & 0 & 0 & 0 & 0 & 0 & 0 & 0 & 0 & 0 & 0 \\
 0 & 0 & 0 & 0 & 0 & 0 & 0 & 0 & 0 & 0 & 0 & 0 & 0 & 0 & 0 & 0 & 0 & 0 & 0 & 0 & 0 & 0 & 0 & 0 & 0 & 0 & 0 & 0 & 0 & 0 & 0 & 0 & 0 & 0 & 0 & 0 & 0 & 0 & 0 & 0 & 0 & 0 \\
 0 & 0 & 0 & 0 & 0 & 0 & 0 & 0 & 0 & 0 & 0 & 0 & 0 & 0 & 0 & 0 & 0 & 0 & 0 & 0 & 0 & 0 & 0 & 0 & 0 & 0 & 0 & 0 & 0 & 0 & 0 & 0 & 0 & 0 & 0 & 0 & 0 & 0 & 0 & 0 & 0 & 0 \\
 0 & 0 & 0 & 0 & 0 & 0 & 0 & 0 & 0 & 0 & 0 & 0 & 0 & 0 & 0 & 0 & 0 & 0 & 0 & 0 & 0 & 0 & 0 & 0 & 0 & 0 & 0 & 0 & 0 & 0 & 0 & 0 & 0 & 0 & 0 & 0 & 0 & 0 & 0 & 0 & 0 & 0 \\
 0 & 0 & 0 & 0 & 0 & 0 & 0 & \frac{3}{2} & -\frac{3}{2} & 0 & 0 & 0 & 0 & 0 & 0 & 0 & 0 & 0 & 0 & 0 & 0 & 0 & 0 & 0 & 0 & 0 & 0 & 0 & 0 & 0 & 0 & 0 & 0 & 0 & 0 & 0 & 0 & 0 & 0 & 0 & 0 & 0 \\
\end{array}
\right)}
 \end{align}

%% file: matrices/3loopM4a.tex
\begin{align}
M_4^{(a)} = \scalemath {0.45}{\left(
\begin{array}{ccccccccccccccccccccccccccccccccccccccccccc}
 0 & 0 & 0 & 0 & 0 & 0 & 0 & 0 & 0 & 0 & 0 & 0 & 0 & 0 & 0 & 0 & 0 & 0 & 0 & 0 & 0 & 0 & 0 & 0 & 0 & 0 & 0 & 0 & 0 & 0 & 0 & 0 & 0 & 0 & 0 & 0 & 0 & 0 & 0 & 0 & 0 & 0 & 0 \\
 0 & 0 & 0 & 0 & 0 & 0 & 0 & 0 & 0 & 0 & 0 & 0 & 0 & 0 & 0 & 0 & 0 & 0 & 0 & 0 & 0 & 0 & 0 & 0 & 0 & 0 & 0 & 0 & 0 & 0 & 0 & 0 & 0 & 0 & 0 & 0 & 0 & 0 & 0 & 0 & 0 & 0 & 0 \\
 0 & 0 & 0 & 0 & 0 & 0 & 0 & 0 & 0 & 0 & 0 & 0 & 0 & 0 & 0 & 0 & 0 & 0 & 0 & 0 & 0 & 0 & 0 & 0 & 0 & 0 & 0 & 0 & 0 & 0 & 0 & 0 & 0 & 0 & 0 & 0 & 0 & 0 & 0 & 0 & 0 & 0 & 0 \\
 0 & 0 & 0 & 0 & 0 & 0 & 0 & 0 & 0 & 0 & 0 & 0 & 0 & 0 & 0 & 0 & 0 & 0 & 0 & 0 & 0 & 0 & 0 & 0 & 0 & 0 & 0 & 0 & 0 & 0 & 0 & 0 & 0 & 0 & 0 & 0 & 0 & 0 & 0 & 0 & 0 & 0 & 0 \\
 0 & 0 & 0 & 0 & 0 & 0 & 0 & 0 & 0 & 0 & 0 & 0 & 0 & 0 & 0 & 0 & 0 & 0 & 0 & 0 & 0 & 0 & 0 & 0 & 0 & 0 & 0 & 0 & 0 & 0 & 0 & 0 & 0 & 0 & 0 & 0 & 0 & 0 & 0 & 0 & 0 & 0 & 0 \\
 0 & 0 & 0 & 0 & 0 & 0 & 0 & 0 & 0 & 0 & 0 & 0 & 0 & 0 & 0 & 0 & 0 & 0 & 0 & 0 & 0 & 0 & 0 & 0 & 0 & 0 & 0 & 0 & 0 & 0 & 0 & 0 & 0 & 0 & 0 & 0 & 0 & 0 & 0 & 0 & 0 & 0 & 0 \\
 0 & 0 & 0 & 0 & 0 & 0 & 0 & 0 & 0 & 0 & 0 & 0 & 0 & 0 & 0 & 0 & 0 & 0 & 0 & 0 & 0 & 0 & 0 & 0 & 0 & 0 & 0 & 0 & 0 & 0 & 0 & 0 & 0 & 0 & 0 & 0 & 0 & 0 & 0 & 0 & 0 & 0 & 0 \\
 0 & 0 & 0 & 0 & 0 & 0 & 0 & 0 & 0 & 0 & 0 & 0 & 0 & 0 & 0 & 0 & 0 & 0 & 0 & 0 & 0 & 0 & 0 & 0 & 0 & 0 & 0 & 0 & 0 & 0 & 0 & 0 & 0 & 0 & 0 & 0 & 0 & 0 & 0 & 0 & 0 & 0 & 0 \\
\end{array}
\right)}
 \end{align}

%% file: matrices/3loopM4b.tex
\begin{align}
M_4^{(b)} = \scalemath {0.45}{\left(
\begin{array}{ccccccccccccccccccccccccccccccccccccccccccc}
 0 & -\frac{4}{3} & \frac{4}{3} & 0 & 0 & 0 & 0 & 0 & 0 & 0 & 0 & 0 & 0 & 0 & 0 & 0 & 0 & 0 & 0 & 0 & 0 & 0 & 0 & 0 & 0 & 0 & 0 & 0 & 0 & 0 & 0 & 0 & 0 & 0 & 0 & 0 & 0 & 0 & 0 & 0 & 0 & 0 & 0 \\
 0 & 0 & 0 & 0 & 0 & 0 & 0 & 0 & 0 & 0 & 0 & 0 & 0 & 0 & 0 & 0 & 0 & 0 & 0 & 0 & 0 & 0 & 0 & 0 & 0 & 0 & 0 & 0 & 0 & 0 & 0 & 0 & 0 & 0 & 0 & 0 & 0 & 0 & 0 & 0 & 0 & 0 & 0 \\
 0 & 0 & 0 & 0 & 0 & 0 & 0 & 0 & 0 & 0 & 0 & 0 & 0 & 0 & 0 & 0 & 0 & 0 & 0 & 0 & 0 & 0 & 0 & 0 & 0 & 0 & 0 & 0 & 0 & 0 & 0 & 0 & 0 & 0 & 0 & 0 & 0 & 0 & 0 & 0 & 0 & 0 & 0 \\
 0 & 0 & 0 & 0 & 0 & 0 & 0 & 0 & 0 & 0 & 0 & 0 & 0 & 0 & 0 & 0 & 0 & 0 & 0 & 0 & 0 & 0 & 0 & 0 & 0 & 0 & 0 & 0 & 0 & 0 & 0 & 0 & 0 & 0 & 0 & 0 & 0 & 0 & 0 & 0 & 0 & 0 & 0 \\
 0 & 0 & 0 & 0 & 0 & 0 & 0 & 0 & 0 & 0 & 0 & 0 & 0 & 0 & 0 & 0 & 0 & 0 & 0 & 0 & 0 & 0 & 0 & 0 & 0 & 0 & 0 & 0 & 0 & 0 & 0 & 0 & 0 & 0 & 0 & 0 & 0 & 0 & 0 & 0 & 0 & 0 & 0 \\
 0 & 0 & 0 & 0 & 0 & 0 & 0 & 0 & 0 & 0 & 0 & 0 & 0 & 0 & 0 & 0 & 0 & 0 & 0 & 0 & 0 & 0 & 0 & 0 & 0 & 0 & 0 & 0 & 0 & 0 & 0 & 0 & 0 & 0 & 0 & 0 & 0 & 0 & 0 & 0 & 0 & 0 & 0 \\
 0 & 0 & 0 & 0 & 0 & 0 & 0 & 0 & 0 & 0 & 0 & 0 & 0 & 0 & 0 & 0 & 0 & 0 & 0 & 0 & 0 & 0 & 0 & 0 & 0 & 0 & 0 & 0 & 0 & 0 & 0 & 0 & 0 & 0 & 0 & 0 & 0 & 0 & 0 & 0 & 0 & 0 & 0 \\
 0 & 0 & 0 & 0 & 0 & 0 & 0 & 0 & 0 & 0 & 0 & 0 & 0 & 0 & 0 & 0 & 0 & 0 & 0 & 0 & 0 & 0 & 0 & 0 & 0 & 0 & 0 & 0 & 0 & 0 & 0 & 0 & 0 & 0 & 0 & 0 & 0 & 0 & 0 & 0 & 0 & 0 & 0 \\
\end{array}
\right)}
 \end{align}

%% file: matrices/3loopM4c.tex
\begin{align}
M_4^{(c)} = \scalemath {0.45}{\left(
\begin{array}{cccccccccccccccccccccccccccccccccccccccccc}
 0 & 0 & 0 & 0 & 0 & 0 & 0 & 0 & 0 & 0 & 0 & 0 & 0 & 0 & 0 & 0 & 0 & 0 & 0 & 0 & 0 & 0 & 0 & 0 & 0 & 0 & 0 & 0 & 0 & 0 & 0 & 0 & 0 & 0 & 0 & 0 & 0 & 0 & 0 & 0 & 0 & 0 \\
 0 & 0 & 0 & 0 & 0 & 0 & 0 & 0 & 0 & 0 & 0 & 0 & 0 & 0 & 0 & 0 & 0 & 0 & 0 & 0 & 0 & 0 & 0 & 0 & 0 & 0 & 0 & 0 & 0 & 0 & 0 & 0 & 0 & 0 & 0 & 0 & 0 & 0 & 0 & 0 & 0 & 0 \\
 0 & 0 & 0 & 0 & 0 & 0 & 0 & 0 & 0 & 0 & 0 & 0 & 0 & 0 & 0 & 0 & 0 & 0 & 0 & 0 & 0 & 0 & 0 & 0 & 0 & 0 & 0 & 0 & 0 & 0 & 0 & 0 & 0 & 0 & 0 & 0 & 0 & 0 & 0 & 0 & 0 & 0 \\
 0 & 0 & 0 & 0 & 0 & 0 & 0 & 0 & 0 & 0 & 0 & 0 & 0 & 0 & 0 & 0 & 0 & 0 & 0 & 0 & 0 & 0 & 0 & 0 & 0 & 0 & 0 & 0 & 0 & 0 & 0 & 0 & 0 & 0 & 0 & 0 & 0 & 0 & 0 & 0 & 0 & 0 \\
 0 & 0 & 0 & 0 & 0 & 0 & 0 & 0 & 0 & 0 & 0 & 0 & 0 & 0 & 0 & 0 & 0 & 0 & 0 & 0 & 0 & 0 & 0 & 0 & 0 & 0 & 0 & 0 & 0 & 0 & 0 & 0 & 0 & 0 & 0 & 0 & 0 & 0 & 0 & 0 & 0 & 0 \\
 0 & 0 & 0 & 0 & 0 & 0 & 0 & 0 & 0 & 0 & 0 & 0 & 0 & 0 & 0 & 0 & 0 & 0 & 0 & 0 & 0 & 0 & 0 & 0 & 0 & 0 & 0 & 0 & 0 & 0 & 0 & 0 & 0 & 0 & 0 & 0 & 0 & 0 & 0 & 0 & 0 & 0 \\
 0 & 0 & 0 & 0 & 0 & 0 & 0 & 0 & 0 & 0 & 0 & 0 & 0 & 0 & 0 & 0 & 0 & 0 & 0 & 0 & 0 & 0 & 0 & 0 & 0 & 0 & 0 & 0 & 0 & 0 & 0 & 0 & 0 & 0 & 0 & 0 & 0 & 0 & 0 & 0 & 0 & 0 \\
 0 & 0 & 0 & 0 & 0 & 0 & 0 & -4 & 0 & 0 & 0 & 0 & 0 & 0 & 0 & 0 & 0 & 0 & 0 & 0 & 0 & 0 & 0 & 0 & 0 & 0 & 0 & 0 & 0 & 0 & 0 & 0 & 0 & 0 & 0 & 0 & 0 & 0 & 0 & 0 & 0 & 0 \\
\end{array}
\right)}
 \end{align}

%% file: matrices/3loopM5.tex
\begin{small}
$(M_5)_{59,58}=-12$, $(M_5)_{68,42}=(M_5)_{68,58}=-2$, $(M_5)_{58,42}=-1$, $(M_5)_{58,40}=(M_5)_{60,42}=(M_5)_{68,40}=1$, $(M_5)_{74,74}=(M_5)_{81,74}=2$, $(M_5)_{33,33}=(M_5)_{58,58}=4$, $(M_5)_{42,42}=5$, $(M_5)_{40,40}=(M_5)_{59,40}=(M_5)_{75,74}=(M_5)_{76,74}=6$, $(M_5)_{35,33}=8$, $(M_5)_{59,42}=12$.
 \end{small}

%% file: matrices/3loopM6a.tex
\begin{align}
M_6^{(a)} = \scalemath {0.45}{\left(
\begin{array}{ccccccccccccccccccccccccccccccccccccccccccc}
 0 & 0 & 0 & 0 & 0 & 0 & 0 & 0 & 0 & 0 & 0 & 0 & 0 & 0 & 0 & 0 & 0 & 0 & 0 & 0 & 0 & 0 & 0 & 0 & 0 & 0 & 0 & 0 & 0 & 0 & 0 & 0 & 0 & 0 & 0 & 0 & 0 & 0 & 0 & 0 & 0 & 0 & 0 \\
 0 & 0 & 0 & 0 & 0 & 0 & 0 & 0 & 0 & 0 & 0 & 0 & 0 & 0 & 0 & 0 & 0 & 0 & 0 & 0 & 0 & 0 & 0 & 0 & 0 & 0 & 0 & 0 & 0 & 0 & 0 & 0 & 0 & 0 & 0 & 0 & 0 & 0 & 0 & 0 & 0 & 0 & 0 \\
 0 & 0 & 0 & 0 & 0 & 0 & 0 & 0 & 0 & 0 & 0 & 0 & 0 & 0 & 0 & 0 & 0 & 0 & 0 & 0 & 0 & 0 & 0 & 0 & 0 & 0 & 0 & 0 & 0 & 0 & 0 & 0 & 0 & 0 & 0 & 0 & 0 & 0 & 0 & 0 & 0 & 0 & 0 \\
 0 & 0 & 0 & 0 & 0 & 0 & 0 & 0 & 0 & 0 & 0 & 0 & 0 & 0 & 0 & 0 & 0 & 0 & 0 & 0 & 0 & 0 & 0 & 0 & 0 & 0 & 0 & 0 & 0 & 0 & 0 & 0 & 0 & 0 & 0 & 0 & 0 & 0 & 0 & 0 & 0 & 0 & 0 \\
 0 & 0 & 0 & 0 & 0 & 0 & 0 & 0 & 0 & 0 & 0 & 0 & 0 & 0 & 0 & 0 & 0 & 0 & 0 & 0 & 0 & 0 & 0 & 0 & 0 & 0 & 0 & 0 & 0 & 0 & 0 & 0 & 0 & 0 & 0 & 0 & 0 & 0 & 0 & 0 & 0 & 0 & 0 \\
 0 & 0 & 0 & 0 & 0 & 0 & 0 & 0 & 0 & 0 & 0 & 0 & 0 & 0 & 0 & 0 & 0 & 0 & 0 & 0 & 0 & 0 & 0 & 0 & 0 & 0 & 0 & 0 & 0 & 0 & 0 & 0 & 0 & 0 & 0 & 0 & 0 & 0 & 0 & 0 & 0 & 0 & 0 \\
 0 & 0 & 0 & 0 & 0 & 0 & 0 & 0 & 0 & 0 & 0 & 0 & 0 & 0 & 0 & 0 & 0 & 0 & 0 & 0 & 0 & 0 & 0 & 0 & 0 & 0 & 0 & 0 & 0 & 0 & 0 & 0 & 0 & 0 & 0 & 0 & 0 & 0 & 0 & 0 & 0 & 0 & 0 \\
 0 & 0 & 0 & 0 & 0 & 0 & 0 & 0 & 0 & 0 & 0 & 0 & 0 & 0 & 0 & 0 & 0 & 0 & 0 & 0 & 0 & 0 & 0 & 0 & 0 & 0 & 0 & 0 & 0 & 0 & 0 & 0 & 0 & 0 & 0 & 0 & 0 & 0 & 0 & 0 & 0 & 0 & 0 \\
\end{array}
\right)}
 \end{align}

%% file: matrices/3loopM6b.tex
\begin{align}
M_6^{(b)} = \scalemath {0.45}{\left(
\begin{array}{ccccccccccccccccccccccccccccccccccccccccccc}
 0 & \frac{4}{3} & 0 & 0 & 0 & 0 & 0 & 0 & 0 & 0 & 0 & 0 & 0 & 0 & 0 & 4 & 0 & 0 & 0 & 0 & 0 & 0 & 0 & 0 & 0 & 0 & 0 & 0 & 0 & 0 & 0 & 0 & 0 & 0 & 0 & 0 & 0 & 0 & 0 & 0 & 0 & 0 & 0 \\
 0 & 0 & 0 & 0 & 0 & 0 & 0 & 0 & 0 & 0 & 0 & 0 & 0 & 0 & 0 & 0 & 0 & 0 & 0 & 0 & 0 & 0 & 0 & 0 & 0 & 0 & 0 & 0 & 0 & 0 & 0 & 0 & 0 & 0 & 0 & 0 & 0 & 0 & 0 & 0 & 0 & 0 & 0 \\
 0 & 0 & 0 & 0 & 0 & 0 & 0 & 0 & 0 & 0 & 0 & 0 & 0 & 0 & 0 & 0 & 0 & 0 & 0 & 0 & 0 & 0 & 0 & 0 & 0 & 0 & 0 & 0 & 0 & 0 & 0 & 0 & 0 & 0 & 0 & 0 & 0 & 0 & 0 & 0 & 0 & 0 & 0 \\
 0 & 0 & 0 & 0 & 0 & 0 & 0 & 0 & 0 & 0 & 0 & 0 & 0 & 0 & 0 & 0 & 0 & 0 & 0 & 0 & 0 & 0 & 0 & 0 & 0 & 0 & 0 & 0 & 0 & 0 & 0 & 0 & 0 & 0 & 0 & 0 & 0 & 0 & 0 & 0 & 0 & 0 & 0 \\
 0 & 0 & 0 & 0 & 0 & 0 & 0 & 0 & 0 & 0 & 0 & 0 & 0 & 0 & 0 & 0 & 0 & 0 & 0 & 0 & 0 & 0 & 0 & 0 & 0 & 0 & 0 & 0 & 0 & 0 & 0 & 0 & 0 & 0 & 0 & 0 & 0 & 0 & 0 & 0 & 0 & 0 & 0 \\
 0 & 0 & 0 & 0 & 0 & 0 & 0 & 0 & 0 & 0 & 0 & 0 & 0 & 0 & 0 & 0 & 0 & 0 & 0 & 0 & 0 & 0 & 0 & 0 & 0 & 0 & 0 & 0 & 0 & 0 & 0 & 0 & 0 & 0 & 0 & 0 & 0 & 0 & 0 & 0 & 0 & 0 & 0 \\
 0 & 0 & 0 & 0 & 0 & 0 & 0 & 0 & 0 & 0 & 0 & 0 & 0 & 0 & 0 & 0 & 0 & 0 & 0 & 0 & 0 & 0 & 0 & 0 & 0 & 0 & 0 & 0 & 0 & 0 & 0 & 0 & 0 & 0 & 0 & 0 & 0 & 0 & 0 & 0 & 0 & 0 & 0 \\
 0 & 0 & 0 & 0 & 0 & 0 & 0 & \frac{7}{2} & 0 & 0 & 0 & 0 & -\frac{3}{4} & 0 & 0 & 0 & \frac{1}{2} & 0 & 0 & 0 & 0 & 0 & 0 & 0 & 0 & 0 & 0 & 0 & 0 & \frac{3}{2} & 0 & 0 & \frac{3}{2} & -\frac{1}{2} & -\frac{9}{4} & \frac{1}{2} & \frac{5}{2} & 0 & 0 & 0 & 0 & 0 & 0 \\
\end{array}
\right)}
 \end{align}

%% file: matrices/3loopM6c.tex
\begin{align}
M_6^{(c)} = \scalemath {0.45}{\left(
\begin{array}{cccccccccccccccccccccccccccccccccccccccccc}
 1 & 0 & 0 & 0 & 0 & 0 & 0 & 0 & 0 & 0 & 0 & 0 & 0 & 0 & 0 & 0 & 0 & 0 & 0 & 0 & 0 & 0 & 0 & 0 & 0 & 0 & 0 & 0 & 0 & 0 & 0 & 0 & 0 & 0 & 0 & 0 & 0 & 0 & 0 & 0 & 0 & 0 \\
 0 & 0 & 0 & 0 & 0 & 0 & 0 & 0 & 0 & 0 & 0 & 0 & 0 & 0 & 0 & 0 & 0 & 0 & 0 & 0 & 0 & 0 & 0 & 0 & 0 & 0 & 0 & 0 & 0 & 0 & 0 & 0 & 0 & 0 & 0 & 0 & 0 & 0 & 0 & 0 & 0 & 0 \\
 0 & 0 & 0 & 0 & 0 & 0 & 0 & 0 & 0 & 0 & 0 & 0 & 0 & 0 & 0 & 0 & 0 & 0 & 0 & 0 & 0 & 0 & 0 & 0 & 0 & 0 & 0 & 0 & 0 & 0 & 0 & 0 & 0 & 0 & 0 & 0 & 0 & 0 & 0 & 0 & 0 & 0 \\
 0 & 0 & 0 & 0 & 0 & 0 & 0 & 0 & 0 & 0 & 0 & 0 & 0 & 0 & 0 & 0 & 0 & 0 & 0 & 0 & 0 & 0 & 0 & 0 & 0 & 0 & 0 & 0 & 0 & 0 & 0 & 0 & 0 & 0 & 0 & 0 & 0 & 0 & 0 & 0 & 0 & 0 \\
 0 & 0 & 0 & 0 & 0 & 0 & 0 & 0 & 0 & 0 & 0 & 0 & 0 & 0 & 0 & 0 & 0 & 0 & 0 & 0 & 0 & 0 & 0 & 0 & 0 & 0 & 0 & 0 & 0 & 0 & 0 & 0 & 0 & 0 & 0 & 0 & 0 & 0 & 0 & 0 & 0 & 0 \\
 0 & 0 & 0 & 0 & 0 & 0 & 0 & 0 & 0 & 0 & 0 & 0 & 0 & 0 & 0 & 0 & 0 & 0 & 0 & 0 & 0 & 0 & 0 & 0 & 0 & 0 & 0 & 0 & 0 & 0 & 0 & 0 & 0 & 0 & 0 & 0 & 0 & 0 & 0 & 0 & 0 & 0 \\
 0 & 0 & 0 & 0 & 0 & 0 & 0 & 0 & 0 & 0 & 0 & 0 & 0 & 0 & 0 & 0 & 0 & 0 & 0 & 0 & 0 & 0 & 0 & 0 & 0 & 0 & 0 & 0 & 0 & 0 & 0 & 0 & 0 & 0 & 0 & 0 & 0 & 0 & 0 & 0 & 0 & 0 \\
 0 & 0 & 0 & 0 & 0 & 0 & 0 & \frac{3}{2} & \frac{3}{2} & 0 & 0 & 0 & 0 & 0 & 0 & 0 & 0 & 0 & 0 & 0 & 0 & 0 & 0 & 0 & 0 & 0 & 0 & 0 & 0 & 0 & 0 & 0 & 0 & 0 & 0 & 0 & 0 & 0 & 0 & 0 & 0 & 0 \\
\end{array}
\right)}
 \end{align}

%% file: 3loopg83.tex
\section{Three-Loop scalar ladder}
\label{App:3loopg83}
As an example, we provide the explicit result for the scalar ladder in~\eqref{def:3loopCanonicalMIs},

\begin{align}
g_{83} &=  \sum_{n=0}^6 \eps^n g_{83}^{(n)} + \mathcal{O}(\eps^7)\,.
\end{align}
In order to reduce the clutter, we use here the notation 

\begin{align}
G_{w_1,\ldots,w_n}(x) &\equiv G(w_1,\ldots,w_n;x)\,.
\end{align}
The coefficients of the $\eps$-expansion of $g_{83}$ are:
{\allowdisplaybreaks[1]
\begin{subequations}
\begin{small}
%
\end{small}
\end{subequations}
}